\documentclass[pra,preprint,superscriptaddress,floatfix,showpacs]{revtex4-1}

\usepackage{amsmath,amsfonts,amssymb}
\usepackage{graphicx}

\def\beq{\begin{equation}}
\def\eeq{\end{equation}}
\def\beqn{\begin{eqnarray}}
\def\eeqn{\end{eqnarray}}

\def\r {{\bf r}}

\begin{document}

\title{Variance of an anisotropic Bose-Einstein condensate}
\author{Shachar Klaiman}
\affiliation{Theoretische Chemie, Physikalisch--Chemisches Institut, Universit\"at Heidelberg, 
Im Neuenheimer Feld 229, D-69120 Heidelberg, Germany}
\author{Raphael Beinke}
\affiliation{Theoretische Chemie, Physikalisch--Chemisches Institut, Universit\"at Heidelberg, 
Im Neuenheimer Feld 229, D-69120 Heidelberg, Germany}
\author{Lorenz S. Cederbaum}
\affiliation{Theoretische Chemie, Physikalisch--Chemisches Institut, Universit\"at Heidelberg, 
Im Neuenheimer Feld 229, D-69120 Heidelberg, Germany}
\author{Alexej I. Streltsov}
\affiliation{Theoretische Chemie, Physikalisch--Chemisches Institut, Universit\"at Heidelberg, 
Im Neuenheimer Feld 229, D-69120 Heidelberg, Germany}
\affiliation{Institut f\"ur Physik, Universit\"at Kassel, Heinrich-Plett-Str. 40, 34132 Kassel, Germany}
\author{Ofir E. Alon}
\email{ofir@research.haifa.ac.il}
\affiliation{Department of Physics, University of Haifa at Oranim, Tivon 36006, Israel}
\affiliation{Haifa Research Center for Theoretical Physics and Astrophysics, University of Haifa,
Haifa 3498838, Israel}

\begin{abstract}
The anisotropy of a trap potential can impact the density 
and variance of a Bose-Einstein condensate (BEC) in an opposite manner.
We exemplify this effect
for both the ground state and out-of-equilibrium dynamics
of structureless bosons
interacting by a long-range inter-particle interaction 
and trapped in a two-dimensional single-well potential.
We demonstrate that even when the density of the BEC is, say, 
wider along the $y$ direction and narrower along the $x$ direction, 
its position variance can actually be smaller and momentum variance larger
in the $y$ direction than in the $x$ direction. 
This behavior of the variance in a many-particle system is counterintuitive.
It suggests using the variance as a tool to characterize
the strength of  
correlations along the $y$ and $x$ directions
in a trapped BEC.
\end{abstract}


\maketitle 

\section{Introduction}\label{Intro}

In quantum mechanics, 
the variance of an observable $\hat o(x)$ 
for a particle described by the wave-packet $\psi(x)$
is often used to interpret and quantify the physical behavior of the particle \cite{QM_book}.
For instance, a wider wave-packet has a larger position variance in comparison 
with a narrower wave-packet which has a smaller position variance.
Physically, the two situations are associated, respectively, 
with de-localization and localization of the quantum particle.
In other words, 
we are often used to infer the position (and momentum) variance
from the shape of the density $|\psi(x)|^2$ of the particle.
This intuitive or visual picture may vary in a system made of 
(many) interacting particles.
It is the purpose of this work to demonstrate and investigate 
such many-body effects using Bose-Einstein condensates 
as an instrumental example.  

Bose-Einstein condensates (BECs) made of ultracold atoms have attracted considerable attention \cite{ex1,ex2,ex3,NB1,NB2,rev1,rev2,rev3,rev4,book1,book2,book3}.
On the theory side, 
many of the investigations to describe their properties 
have been performed
using Gross-Pitaevskii (mean-field) theory,
which assumes all bosons to reside in the same orbital.
It is generally perceived 
that Gross-Pitaevskii theory properly describes
the ground state as well as the out-of-equilibrium dynamics of (trapped) BECs in the infinite-particle limit,
when the product of the number of particles times the scattering length 
(i.e., the interaction parameter) is constant.
Indeed, there are rigorous results which
prove (under certain conditions)
that the energy per particle and density per particle of the many-boson system coincide in the infinite-particle limit
with the respective Gross-Piteavekii quantities, 
and that the bosonic system is $100\%$ condensed \cite{Yngvason_PRA,Lieb_PRL,Erdos_PRL,MATH_ERDOS_REF},
also see in this context \cite{Castin_U,L_new}.

Despite the situation that BECs are 100\% in the infinite-particle limit,
and that their density per particle and energy per particle coincide 
with those computed by the Gross-Pitaevskii theory, 
the story does not end here.
In \cite{Variance, TD_Variance} we demonstrated 
that the variance of a many-particle operator and the uncertainty product of
two many-particle operators can substantially deviate from those given by the Gross-Pitaevskii theory,
even in the infinite-particle limit when the bosonic system is 100\% condensed.
Physically, 
these many-body effects are governed by the (often very small) 
number of depleted particles which, unlike the non-condensed fraction,
does not vanish even in the infinite-particle limit.
Mathematically,
the difference between the predictions of the many-body and mean-field descriptions
can be tracked down to the subtlety of 
performing the infinite-particle limit only after (and not before) the many-particle 
operator is evaluated.
In comparison with the variance of an operator of a single particle,
the variance of many-particle operators is a much richer quantity, 
also see \cite{Drumm,Oriol_Robin,BB_HIM,BB_new} in this context.
At the bottom line, 
the many-body and mean-field wave-functions themselves 
differ at the infinite-particle limit and, consequently,
their overlap is always smaller than one 
and can become arbitrarily small \cite{L_new,SL_Psi}.

In the present work we would like to investigate the
intriguing possibilities which open up for the variance of a trapped BEC in two spatial dimensions,
and in particular the connection between shape of the bosonic cloud (i.e., the density)
and its position and momentum variance.
Intermingled with the above is the investigation of the variance along 
the pathway from condensation to fragmentation of a trapped BEC.
The latter is far away from the infinite-particle limit (where the bosonic system is 100\% condensed)
which was the focus of previous work \cite{Variance,TD_Variance}.
We mention that preliminary results in one spatial dimension 
were recently reported in \cite{TD_Variance_BEC}.
The structure of the paper is as follows.
In Sec.~\ref{Theory} we briefly 
discuss the variance in a many-body system 
and its computation from the wave-function of a trapped BEC.
In Sec.~\ref{APPL} we present two detailed investigations,
one of the ground state (Subsec.~\ref{Ex1})
and the second of the out-of-equilibrium dynamics following
an interaction quench (Subsec.~\ref{Ex2}). 
Conclusions are put forward in Sec.~\ref{Conclusions}.
Finally,
further details of the numerics and convergence 
are provided in the Appendix.

\section{Theory}\label{Theory}

We begin with
the many-body Hamiltonian of $N$ interacting bosons
\beq\label{HAM}
 \hat H(\r_1,\ldots,\r_N;\lambda_0) = 
\sum_{j=1}^N \hat h(\r_j) + \sum_{j<k} \lambda_0\hat W(\r_j-\r_k).
\eeq
Here $\hat h(\r) = -\frac{1}{2} \frac{\partial^2}{\partial \r^2} + \hat V(\r)$ is the one-particle Hamiltonian 
where $V(\r)$ the trap potential and $\hat W(\r_1-\r_2)$ the inter-particle interaction of strength $\lambda_0$.
Throughout this work 
$\r=(x,y)$ is the position vector in two spatial dimensions and $\hbar=m=1$.

In the time-independent part
of our work,
$\hat H(\r_1,\ldots,\r_N;\lambda_0) \Phi(\r_1,\ldots,\r_N) =\break\hfill E \Phi(\r_1,\ldots,\r_N)$,
we investigate the ground state of the bosons,
where $E$ is the energy and $\Phi(\r_1,\ldots,\r_N)$ normalized to unity.
In the out-of-equilibrium part
we solve the time-dependent Schr\"odinger equation, 
$\hat H(\r_1,\ldots,\r_N;\lambda'_0) \Psi(\r_1,\ldots,\r_N;t) = i \frac{\partial\Psi(\r_1,\ldots,\r_N;t)}{\partial t}$,
for the scenario where 
the system evolves following an interaction quench from $\lambda_0$ to $\lambda'_0$,
with the initial condition 
$\Psi(\r_1,\ldots,\r_N;0) = \Phi(\r_1,\ldots,\r_N)$.

To analyze the many-body wave-function $\Psi(\r_1,\ldots,\r_N;t)$
we use its reduced one-body and two-body 
density matrices \cite{Lowdin,Yukalov,Mazz,RDMs}.
The reduced one-body density matrix
\beqn\label{1RDM}
\frac{\rho^{(1)}(\r_1,\r_1';t)}{N} &=&
\int d\r_2 \ldots d\r_N \, \Psi^\ast(\r_1',\r_2,\ldots,\r_N;t) \Psi(\r_1,\r_2,\ldots,\r_N;t) = \nonumber \\
 &=& \sum_j \frac{n_j(t)}{N} \, \alpha_j(\r_1;t) \alpha^\ast_j(\r'_1;t)
\eeqn
is prescribed using the natural orbitals $\alpha_j(\r;t)$ and natural occupations $n_j(t)$.
We generally enumerate the occupation numbers in the order of non-increasing values.
We call $\sum_{j>1}n_j(t)$ the number of depleted particles (depletion in short)
and $\frac{\sum_{j>1} n_j(t)}{N}$ the depleted fraction. 
The latter are used to define the degree of condensation  of the interacting bosons \cite{Penrose_Onsager}.
The diagonal of the reduced one-body density matrix,
$\rho(\r;t) = \rho^{(1)}(\r,\r;t)$,
is referred to as the density. 
The diagonal part of the reduced two-body density matrix,
\beqn\label{2RDM}
\frac{\rho^{(2)}(\r_1,\r_2,\r_1,\r_2;t)}{N(N-1)} &=& 
 \int d\r_3 \ldots d\r_N \, \Psi^\ast(\r_1,\r_2,\ldots,\r_N;t) \Psi(\r_1,\r_2,\ldots,\r_N;t) = \nonumber \\
 &=& \sum_{jpkq} \frac{\rho_{jpkq}(t)}{N(N-1)} \, 
\alpha^\ast_j(\r_1;t) \alpha^\ast_p(\r_2;t) \alpha_k(\r_1;t) \alpha_q(\r_2;t), 
\eeqn
is expressed using the natural orbitals,
where
$\rho_{jpkq}(t) = \langle\Psi(t)|\hat b_j^\dag \hat b_p^\dag \hat b_k \hat b_q|\Psi(t)\rangle$
and the creation $\hat b_j$ (and annihilation) operators are associated with $\alpha_j(\r;t)$.

Using the density per particle and natural orbitals, 
the variance per particle of an operator $\hat A=\sum_{j=1}^N \hat a(\r_j)$
which is local in position space
reads \cite{Variance,TD_Variance}
\beqn\label{dis}
& & \frac{1}{N}\Delta_{\hat A}^2(t) = \frac{1}{N} 
\left[\langle\Psi(t)|\hat A^2|\Psi(t)\rangle - \langle\Psi(t)|\hat A|\Psi(t)\rangle^2\right] \equiv 
\Delta_{\hat a, density}^2(t) + \Delta_{\hat a, MB}^2(t), \nonumber \\
& & \quad \Delta_{\hat a, density}^2(t) = 
\int d\r \frac{\rho(\r;t)}{N} a^2(\r) - \left[\int d\r \frac{\rho(\r;t)}{N} a(\r) \right]^2, \nonumber \\ 
& & \quad \Delta_{\hat a, MB}^2(t) = \frac{\rho_{1111}(t)}{N} \left[\int d\r |\alpha_1(\r;t)|^2 a(\r) \right]^2 
- (N-1) \left[\int d\r \frac{\rho(\r;t)}{N} a(\r) \right]^2 + \nonumber \\
& & \quad \quad + \sum_{jpkq\ne 1111} \frac{\rho_{jpkq}(t)}{N} \left[\int d\r \alpha^\ast_j(\r;t) \alpha_k(\r;t) a(\r) \right]
\left[\int d\r \alpha^\ast_p(\r;t) \alpha_q(\r;t) a(\r)\right]. \
\eeqn
The first term, $\Delta_{\hat a, density}^2(t)$,
describes the variance of $\hat a(\r)$ resulting from the density per particle $\frac{\rho(\r;t)}{N}$.
The second term, $\Delta_{\hat a, MB}^2(t)$,
collects all other contributions to the many-particle variance.
$\Delta_{\hat a, MB}^2(t)$ is generally non-zero within a many-body theory,
but is identically equal to
zero within Gross-Pitaevskii theory.
We remark that analogous expressions hold for operators 
which are local in momentum space.

\section{Results}\label{APPL}

Our system of choice is
made of
structureless bosons with harmonic inter-particle interaction trapped
in a single-well {\it anharmonic} potential.
Unlike models of particles interacting with harmonic inter-particle interaction 
and trapped in a harmonic trap, 
which have been extensively
studied and
can be solved analytically \cite{hm1,hm2,hm3,hm4,hm5,hm6,hm7,hm8,hm9,hm10,hm11,hm12,hm13,hm14,hm15},
the Schr\"odinger equation of the trapped BEC
has no analytical solution in the present study,
nor even the variance can be determined analytically,
thus necessitating a numerical treatment.

For this,
we have to employ a suitable many-body theoretical and computational approach.
Such a many-body tool is the multiconfigurational time-dependent
Hartree (MCTDH) for bosons (MCTDHB) method \cite{MCTDHB1,MCTDHB2}
which has been extensively used in the literature 
\cite{BJJ,MCTDHB_OCT,MCTDHB_Shapiro,Iva,LC_NJP,Breaking,Peter_2015a,Peter_2015b,Uwe,Sven_Tom,
Tunneling_Rapha,Peter_b,Alexej_u,Kaspar_n,MCTDHB_spin,Axel_ar,Joachim,Uwe_sub}. 
For further documentation of MCTDHB see \cite{Kaspar_The,book_nick,Axel_The},
and for its benchmarks with an exactly-solvable model \cite{Benchmarks} (and \cite{Axel_MCTDHF_HIM}).
MCTDHB can be seen as the indistinguishable-particle bosonic daughter of MCTDH \cite{cpl1990,jcp1992,review_Dieter,book_MCTDH}.

\subsection{Statics}\label{Ex1}

\begin{figure}[!]
\begin{center}
\includegraphics[width=0.70\columnwidth,angle=-90]{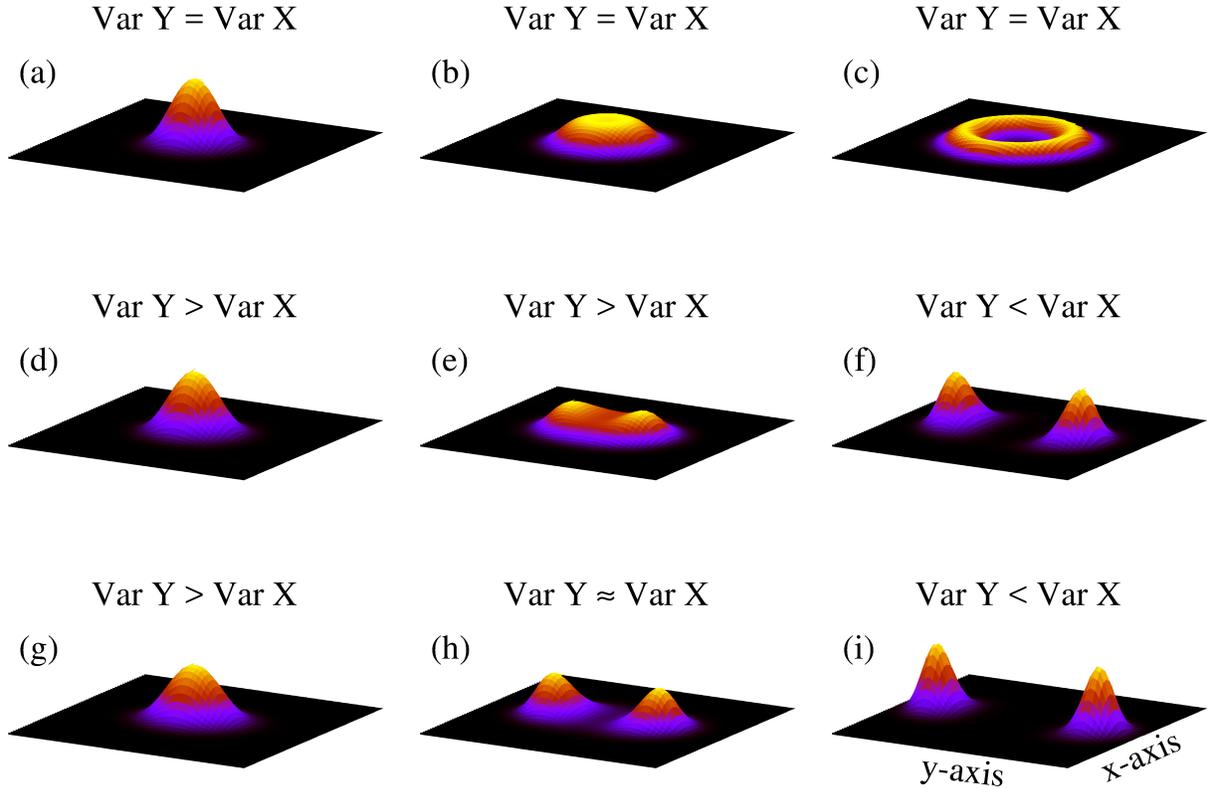}
\end{center}
\caption{(Color online) 
Anisotropy of the variance in the ground state.
Shown is the density per particle 
of $N=10$ bosons for the interaction strengths
$\lambda_0=0.01$, $0.10$, and $0.20$ (left to right columns)
and trap anisotropies $\beta=0\%$ (isotropic), $10\%$, and $20\%$ (upper to lower rows).
Increasing the interaction strength the ground state changes its shape and 
the reduced one-body density matrix fragments, see Fig.~\ref{f2}.
In the upper row, panels (a), (b), and (c), the system is isotropic and naturally the position variances 
in the $y$ and $x$ directions are equal.
In the middle row, panels (d), (e), and (f), and lower row, panels (g), (h), and (i),
the system is anisotropic and, 
as the density in the $y$ direction becomes wider than the density in the $x$ direction,
the respective variances surprisingly 
behave in an opposite manner.
That is,
the variance in the $y$ direction becomes smaller than that in the $x$ direction, see Fig.~\ref{f3}.
This many-body phenomenon, 
where the anisotropy of the position variance behaves in an opposite manner to 
the anisotropy of the density, 
is hence counterintuitive.  
The results are obtained for $M=10$ time-adaptive
(self-consistent) orbitals.
See the text for further discussion.
The quantities shown are dimensionless.}
\label{f1}
\end{figure}

\begin{figure}[!]
\begin{center}
\vglue -2.0 truecm
\hglue -1.0 truecm
\includegraphics[width=0.445\columnwidth,angle=-90]{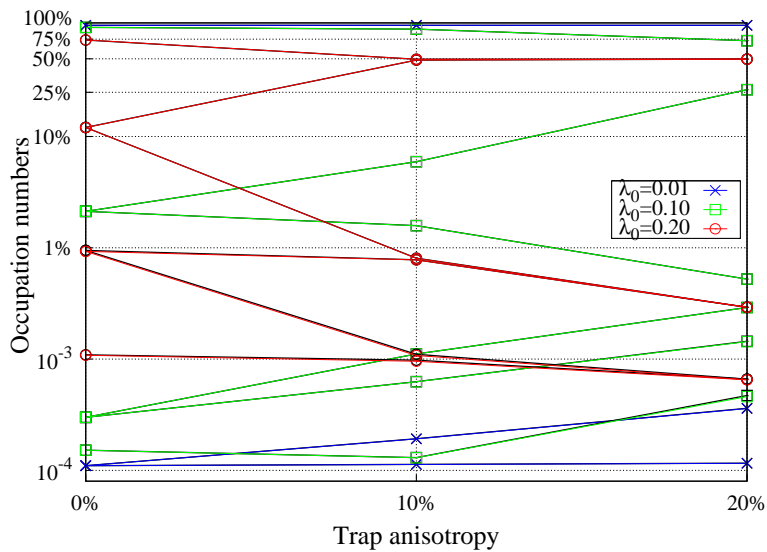}
\end{center}
\vglue 0.25 truecm
\caption{(Color online)
Occupation numbers $n_j/N$ 
for $N=10$ bosons with interactions of different strengths, and in traps of different anisotropies,
throughout the pathway from condensation to fragmentation of the ground state.
The three largest occupation numbers for $\lambda_0=0.01$ (blue curves with stars) computed
with $M=3$ time-adaptive (self-consistent)
orbitals and the six largest occupations numbers for $\lambda_0=0.10$ (green curves with boxes)
and $\lambda_0=0.20$ (red curves with circles) computed with $M=10$ orbitals
(actual data are marked by symbols, the curves are to guide the eye).
Fragmentation of the ground state with increasing interaction strength and anisotropy of the trap
is demonstrated, see the text for further discussion.
Also plotted by 
the black curves with the same palette of symbols 
are the results obtained by including the next `filled shell'
of orbitals, namely, $M=6$ and $M=15$, respectively.
It is seen that the results with $M=3$ and $M=6$ orbitals for the weakest interaction, 
and the results with $M=10$ and $M=15$ orbitals for the stronger interactions lie atop each other.
The quantities shown are dimensionless.}
\label{f2}
\end{figure}

\begin{figure}[!]
\begin{center}
\vglue -2.0 truecm
\hglue -1.0 truecm
\includegraphics[width=0.445\columnwidth,angle=-90]{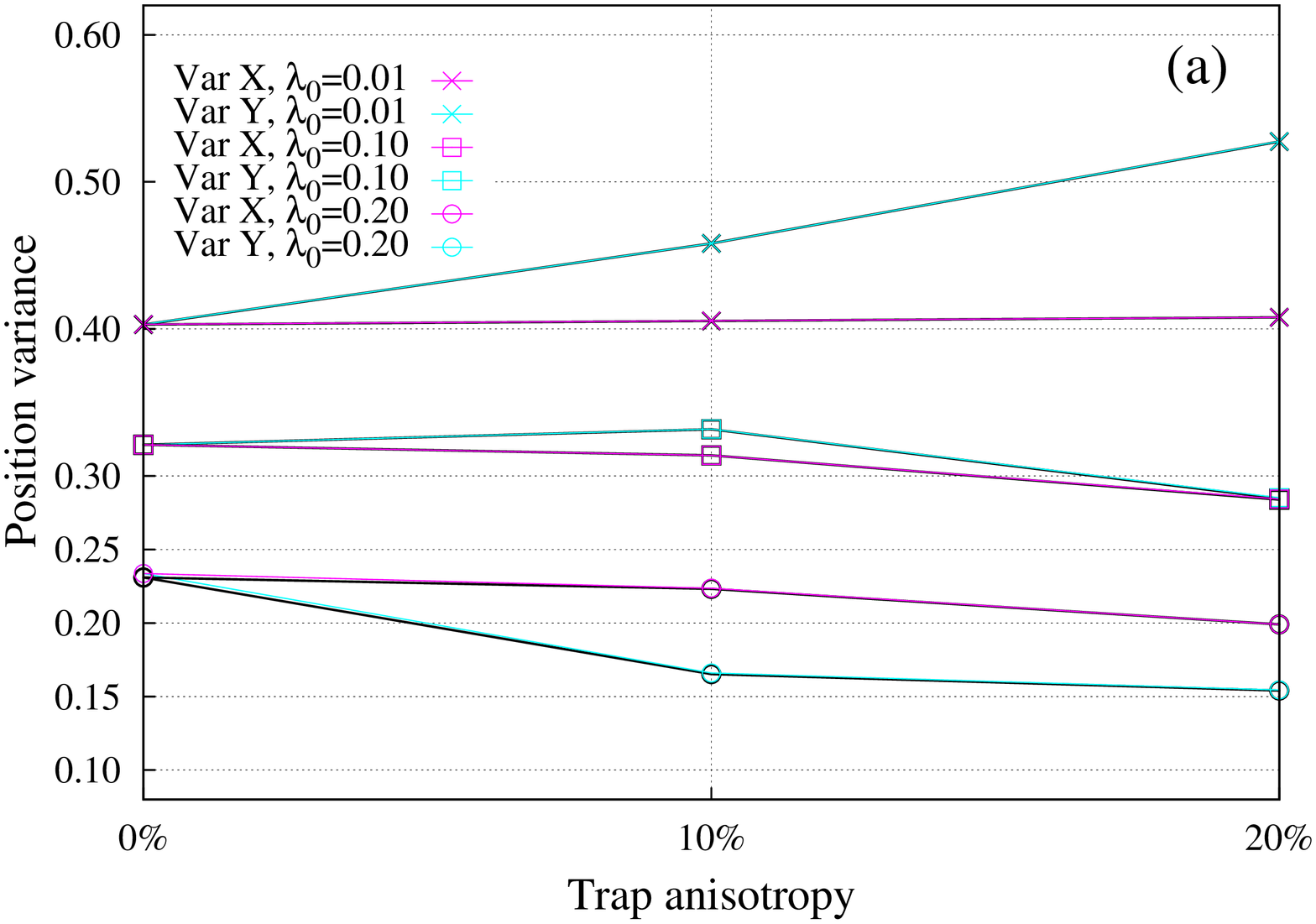}
\hglue -1.0 truecm
\includegraphics[width=0.445\columnwidth,angle=-90]{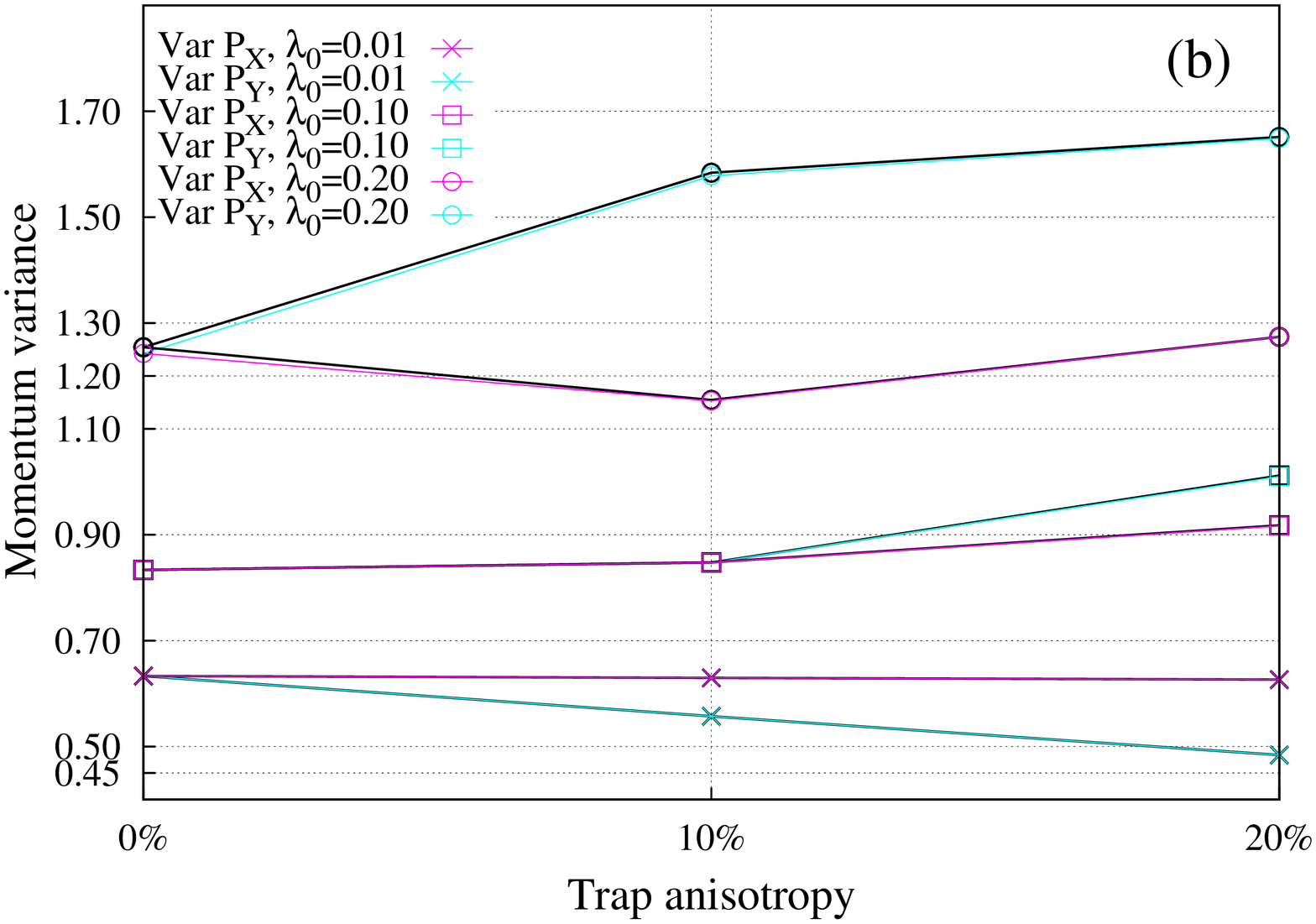}
\end{center}
\vglue 0.25 truecm
\caption{(Color online)
(a) Position variance per particle $\frac{1}{N}\Delta^2_{\hat X}$ and $\frac{1}{N}\Delta^2_{\hat Y}$ 
and (b) momentum variance $\frac{1}{N}\Delta^2_{\hat P_X}$ and $\frac{1}{N}\Delta^2_{\hat P_Y}$ 
for $N=10$ bosons in traps of different anisotropies and for interactions of different strengths
throughout the pathway from condensation to fragmentation of the ground state.
The magenta curves and symbols are for quantities along the $x$ direction 
and the cyan curves and symbols are for quantities along the $y$ direction 
(actual data are marked by symbols, the curves are to guide the eye).
Crosses are data for $\lambda_0=0.01$
(computed with $M=3$ orbitals), 
and boxes are for $\lambda_0=0.10$ and circles for $\lambda_0=0.20$ 
(computed with $M=10$ orbitals).
Anisotropy of the variance 
is demonstrated, see the text and Fig.~\ref{f1} for further discussion.
Also plotted by 
the black curves with the same palette of symbols 
are the results obtained by including the next `filled shell'
of orbitals, namely, $M=6$ and $M=15$, respectively.
It is seen that the results with $M=3$ and $M=6$ orbitals for the weakest interactions, 
and the results with $M=10$ and $M=15$ orbitals for the stronger interactions practically lie atop each other.
The quantities shown are dimensionless.}
\label{f3}
\end{figure}

We investigate the variance along the
pathway from condensation to fragmentation of the ground state of trapped bosons \cite{Sipe,ALN,Pathway,Pathway_Pe,MCHB}.
Fragmentation of BECs has drawn much attention, see, e.g., \cite{frg1,frg2,frg3,frg4,frg5,frg6,frg7,frg8,frg9}.
In particular for structureless bosons with a long-range interaction in a single trap,
the ground state has been shown to become fragmented when increasing the inter-particle repulsion \cite{Uwe,Uwe_PRL1,MCTDHB_3D_stat,Uwe_PRL2,MCTDHB_3D_dyn,Uwe_PRL3}.
Figs.~\ref{f1}, \ref{f2}, and \ref{f3} below
collect the results.

The one-body Hamiltonian in (\ref{HAM}) is 
$\hat h(x,y)=-\left(\frac{1}{2}\frac{\partial^2}{\partial x^2}+\frac{1}{2}\frac{\partial^2}{\partial y^2}\right) +
\frac{\{x^2+[(1-\beta)y]^2\}^2}{4}$,
where the degree of anisotropy is $\beta$.
The inter-particle interaction is harmonic and repulsive, 
$\lambda_0\hat W(x_1-x_2,y_1-y_2)=-\lambda_0[(x_1-x_2)^2+(y_1-y_2)^2], \lambda_0>0$. 
Fig.~\ref{f1} depicts 
snapshots of the density per particle of $N=10$ bosons 
for three interaction strengths $\lambda_0=0.01$, $0.10$, and $0.20$
[the interaction 
parameters are $\Lambda=\lambda_0(N-1)=0.09$, $0.9$, and $1.8$, respectively] 
and three degrees of anisotropy $\beta=0\%$ (isotropic), $10\%$, and $20\%$. 
We follow the changes in the ground state.
The density broadens as the interaction is increased. 
For the isotropic trap,
see the upper row Fig.~\ref{f1}a,b,c,
the density remains, of course, rotationally symmetric and eventually a torus-like shape emerges.
For the anisotropic traps,
the density splits into two clouds,
see the middle row Fig.~\ref{f1}d,e,f 
and lower row Fig.~\ref{f1}g,h,i.
The more anisotropic is the trap, the more split is the ground-state density 
for a given interaction parameter.

Along side the changes in the shape of the ground-state density,
the reduced one-particle density of the ground state fragments \cite{Uwe,Uwe_PRL1,MCTDHB_3D_stat,Uwe_PRL2,MCTDHB_3D_dyn,Uwe_PRL3}.
Depending on the anisotropy of the trap and strength of the interaction,
the ground state evolves from nearly fully condensed, for $\lambda_0=0.01$ and $\beta=0\%$, 
to almost fully two-fold fragmented, for $\lambda_0=0.20$ and $\beta=20\%$,
see Fig.~\ref{f2} and discussion below.
The more anisotropic is the trap, the more fragmented is the ground state for a given interaction parameter.
It is interesting to follow the `correlation diagram' of the occupation numbers [see (\ref{1RDM})] for $\lambda_0=0.20$.
Degenerate occupation numbers for the isotropic trap ($n_2,n_3$ and $n_4,n_5$ for $\beta=0\%$) 
split when the anisotropy sets in,
and together with the non-degenerate occupation numbers ($n_1$ and $n_6$ for $\beta=0\%$) 
essentially merge to pairs ($n_1,n_2$ and $n_3,n_4$ and $n_5,n_6$ for $\beta=20\%$) 
when the fragmentation is full, see Fig.~\ref{f2}.

We now move to the central quantity of interest -- the variance.
Fig.~\ref{f3} depicts the many-particle position variance per particle,
$\frac{1}{N}\Delta^2_{\hat X}$,
and momentum variance,
$\frac{1}{N}\Delta^2_{\hat P}$,
of the ground state for the interaction strengths 
$\lambda_0=0.01$, $0.10$, and $0.20$
and anisotropies of the trap 
$\beta=0\%$, $10\%$, and $20\%$ discussed above.
Generally, enlarging the anisotropy of the trap (in our case along the y-axis) enlarges the area
available for the trapped interacting bosons.
For the weakest interaction, $\lambda_0=0.01$, the position variance along the y-axis increases monotonically,
and that along the x-axis hardly changes, 
see the upper two curves in Fig.~\ref{f3}a.
Side by side, the momentum variance along the y-axis decreases monotonically,
whereas that along the x-axis decreases very mildly,
see the lower two curves in Fig.~\ref{f3}b. 
These are compatible with the shapes of the density, 
see the left column Fig.~\ref{f1}a,d,g,
and with the occupation numbers, see Fig.~\ref{f2},
indicating that the systems are essentially fully condensed
[$\frac{n_1}{N} > 0.99(9)$ for all three anisotropies].

Already for $\lambda_0=0.10$, 
changing the anisotropy of the trap leads to a different behavior of the variance.
At $10\%$ anisotropy,
where the system is somewhat depleted with $\frac{n_1}{N}>0.92$,
the position variance along the y-axis slightly increases and that along the x-axis slightly decreases.
Accidentally, the momentum variance along both directions are approximately equal (and slightly increase).
At $20\%$ anisotropy the position variance along both directions are
incidentally approximately equal (and decrease),
and the momentum variance along the y-axis
is larger than that along the x-axis (both increase).
The system is now already two-fold fragmented with $\frac{n_1}{N}=0.73$ and $\frac{n_2}{N}=0.26$.
Looking at the shapes of the density in the middle column Fig.~\ref{f1}b,e,h,
we observe the 
effects of the depletion and more visibly of the fragmentation on the variance. 
The many-body term of the variance
becomes dominant over the density term of the variance [see (\ref{dis})]
in the ground state
of the system. 

The results for $\lambda_0=0.20$ are even more prominent. 
First, despite the fact that the
density broadens along the y-axis
while the anisotropy is enlarged,
the position variance decreases (in both directions, but more mildly in the $x$ direction).
Moreover and in an opposite manner, 
the y-axis position variance is smaller than that along the x-axis,
see lower two curves in Fig.~\ref{f3}a,
although the density along the y-axis is much broader than that along the x-axis,
see the right column Fig.~\ref{f1}c,f,i.
The behavior of the momentum variance
is in line with the above many-body effects.
The momentum variance along the y-axis is larger than that along the x-axis,
i.e., opposite to the shape of the density. 
Furthermore,
the momentum variance increases monotonously along the $y$ direction,
but first decreases and than increases along the $x$ direction.
The occupation numbers, $\frac{n_1}{N}=0.49(5)$ and $\frac{n_2}{N}=0.49$ for $\beta=10\%$
and $\frac{n_1}{N} \approx 0.5$ and $\frac{n_2}{N}\approx 0.5$ for $\beta=20\%$,
signify that the system becomes essentially fully two-fold fragmented.
Again and more pronouncedly,
the many-body term of the variance is 
dominant over the density term of the variance [see (\ref{dis})]
when the ground state is fragmented.

Let us recapitulate.
As a two-dimensional isotropic anharmonic 
trap is stretched along the $y$ direction,
the density basically
broadens in that direction.
The long-range interaction causes the system to fragment.
Then, the position variance decreases and the momentum variance increases,
unlike from what one could anticipate  
by just examining the shape of the density.
On top of that,
we find that, although the density of the cloud is anisotropic,
here it is broader along the y-axis
than along the x-axis, 
the position variance along the $y$ direction is smaller than that along the $x$ direction.
Correspondingly, the momentum variance along the $y$ direction is larger than that along the $x$ direction.
We stress that
this is contrary to what one would 
expect by just examining the shape of the two-dimensional density.
All in all,
these are intriguing many-body effects 
in the ground state of a fragmented trapped BEC.   

\subsection{Dynamics}\label{Ex2}

The counter-intuitive properties of the position and momentum variances discussed in Subsec.~\ref{Ex1}
are associated with the 
larger depleted fraction and more pronounced 
fragmentation of the ground state
of a finite Bose system.
The natural question to ask is whether such or similar effects 
can occur in larger systems.
In the present subsection we shall demonstrate that the answer is
positive and concentrate on an out-of-equilibrium scenario.
We thereby keep in mind that we are {\it en route} the limit of an infinite number of particles 
where the depletion per particle of the system diminishes to zero. 

The one-body Hamiltonian is again 
$\hat h(x,y)=-\left(\frac{1}{2}\frac{\partial^2}{\partial x^2}+
\frac{1}{2}\frac{\partial^2}{\partial y^2}\right) + \frac{\{x^2+[(1-\beta)y]^2\}^2}{4}$,
where the degree of anisotropy is $\beta=0\%$, $10\%$, and $20\%$,
and the inter-particle interaction is harmonic and repulsive, 
$\lambda_0\hat W(x_1-x_2,y_1-y_2)=-\lambda_0[(x_1-x_2)^2+(y_1-y_2)^2], \lambda_0>0$.
We consider the smallest interaction parameter from the above ground-state study,
$\Lambda = \lambda_0(N-1)=0.09$,
but take a hundred times larger number of interacting bosons, $N=1000$.
The system is prepared in the ground state of the trap.
The initial depletion fraction 
is accordingly a hundred times smaller [$\frac{n_1(0)}{N} > 0.999 \, 9(9)$]
for each of the three anisotropies $\beta=0\%$, $10\%$, and $20\%$
than for the corresponding cases with $N=10$ bosons in Subsec.~\ref{Ex1}.
The initial densities per particle
look just like the left column Fig.~\ref{f1}a,d,g. 

\begin{figure}[!]
\begin{center}
\vglue -2.0 truecm
\hglue -1.0 truecm
\includegraphics[width=0.345\columnwidth,angle=-90]{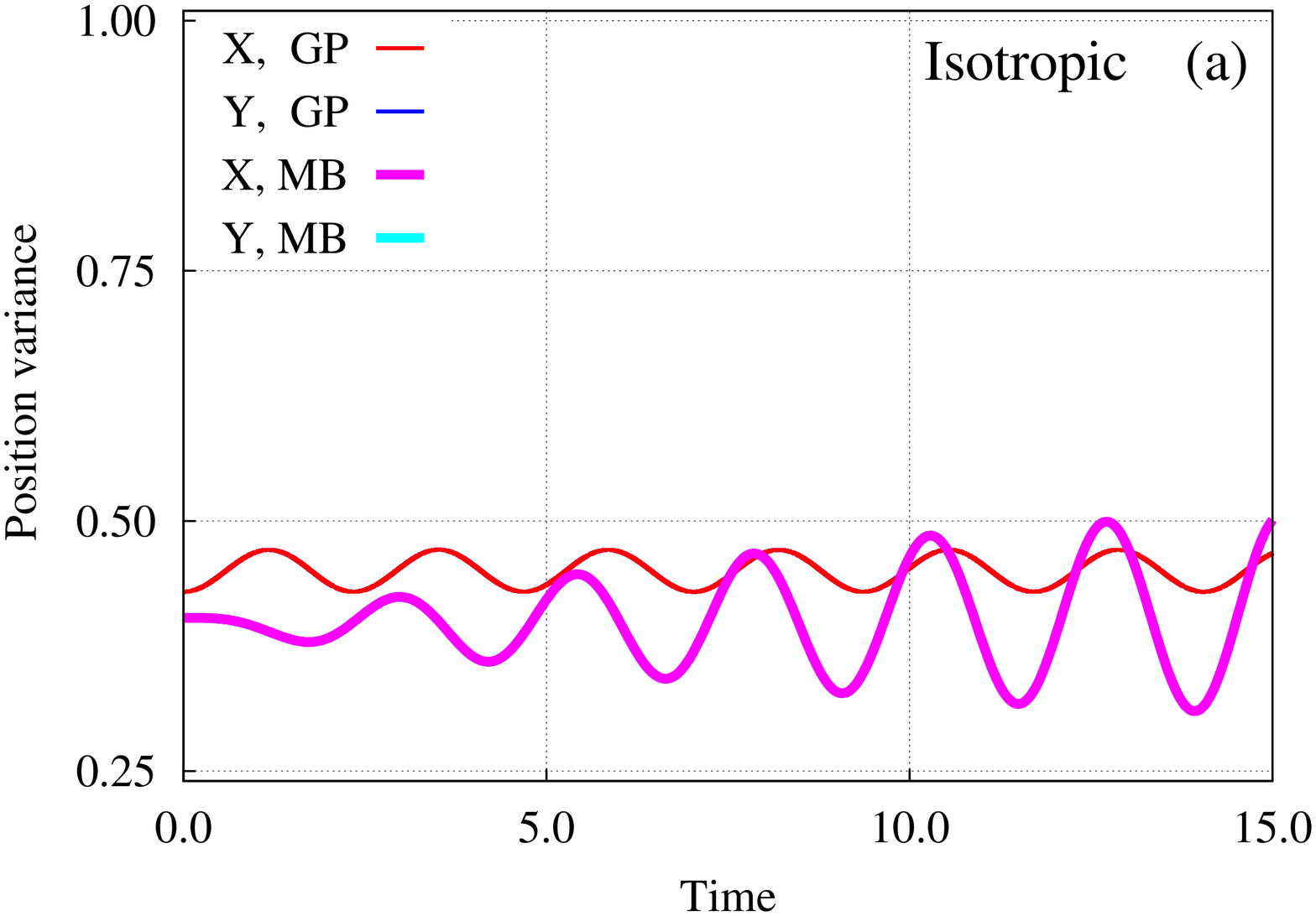}
\includegraphics[width=0.345\columnwidth,angle=-90]{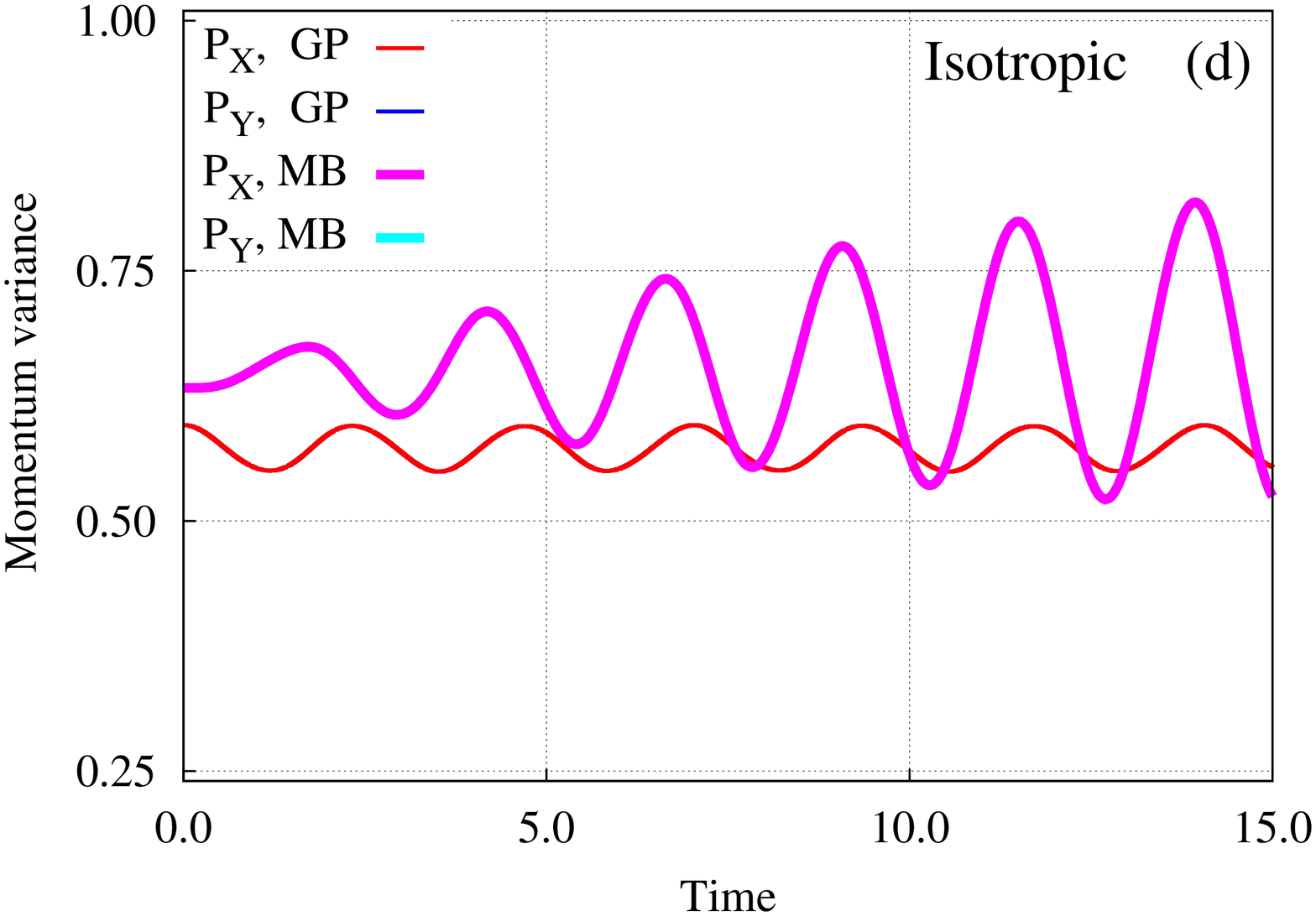}
\hglue -1.0 truecm
\includegraphics[width=0.345\columnwidth,angle=-90]{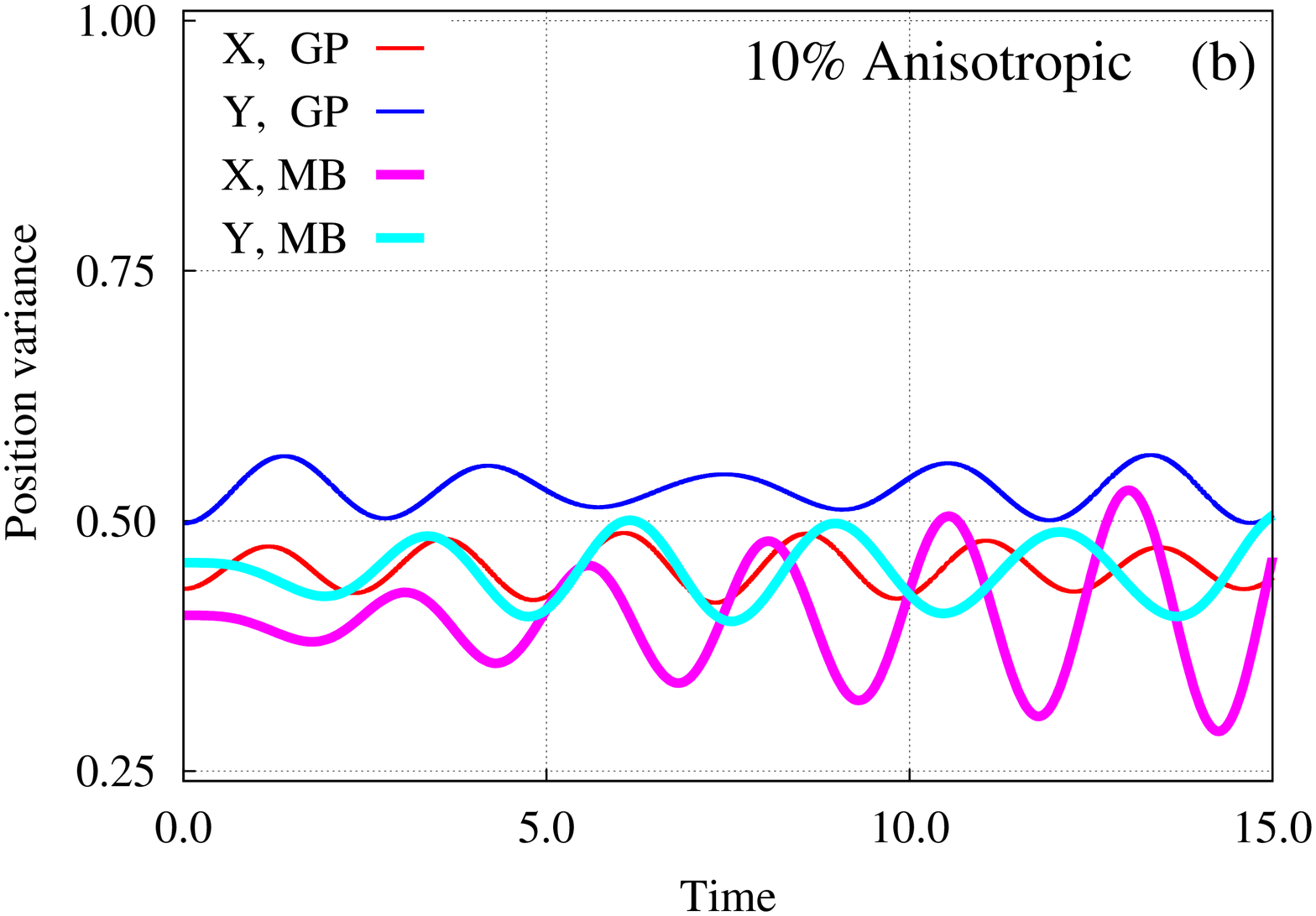}
\includegraphics[width=0.345\columnwidth,angle=-90]{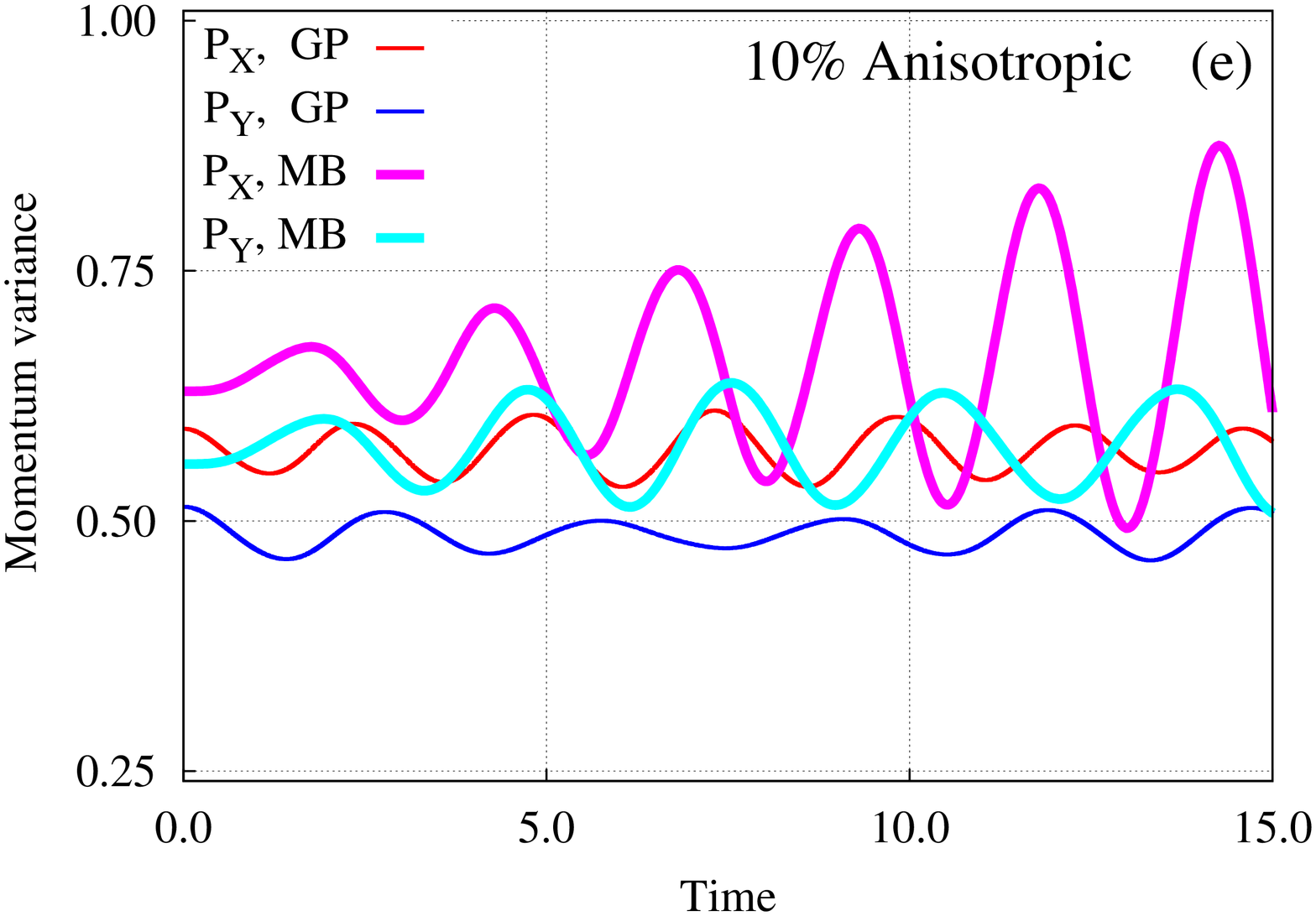}
\hglue -1.0 truecm
\includegraphics[width=0.345\columnwidth,angle=-90]{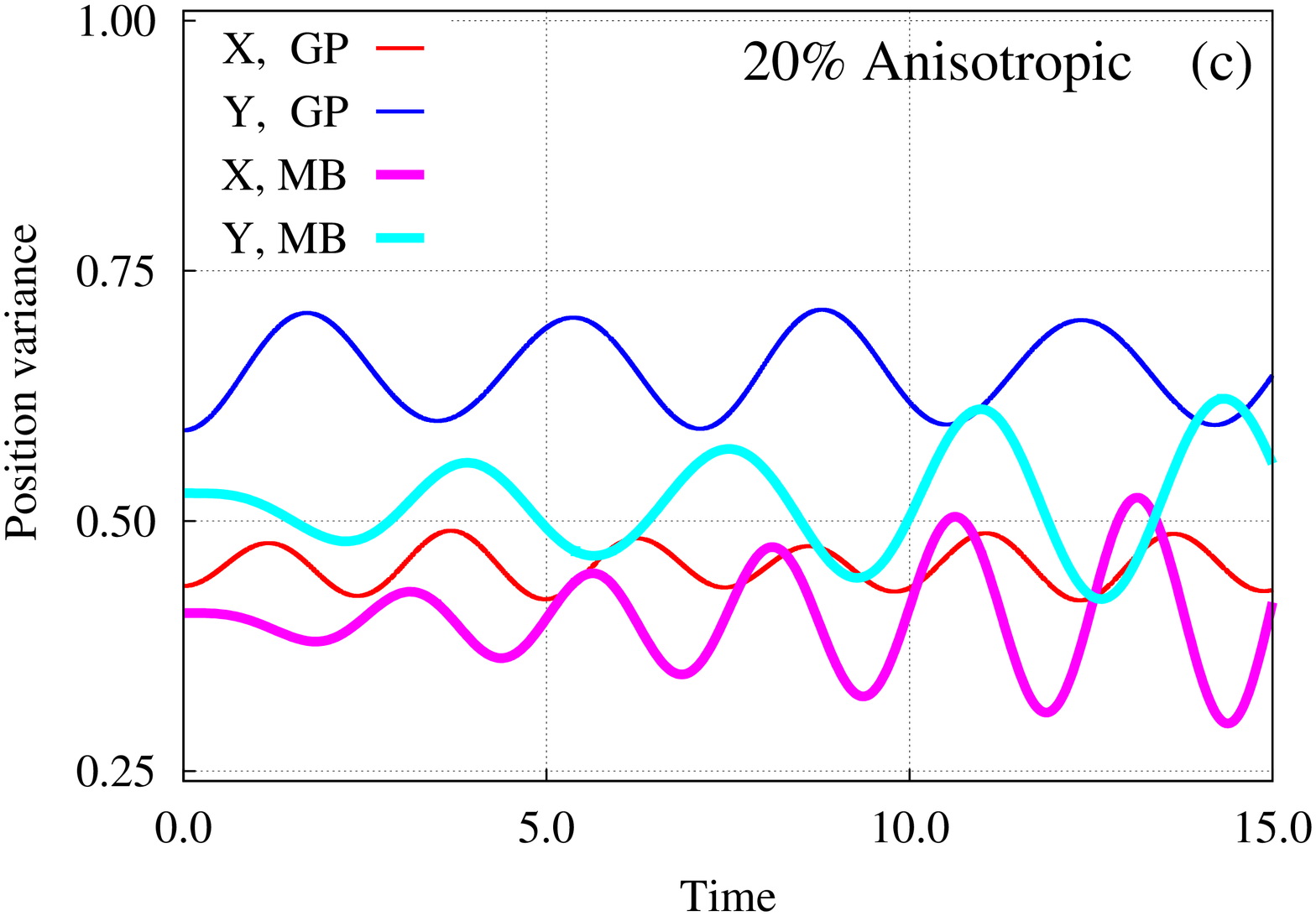}
\includegraphics[width=0.345\columnwidth,angle=-90]{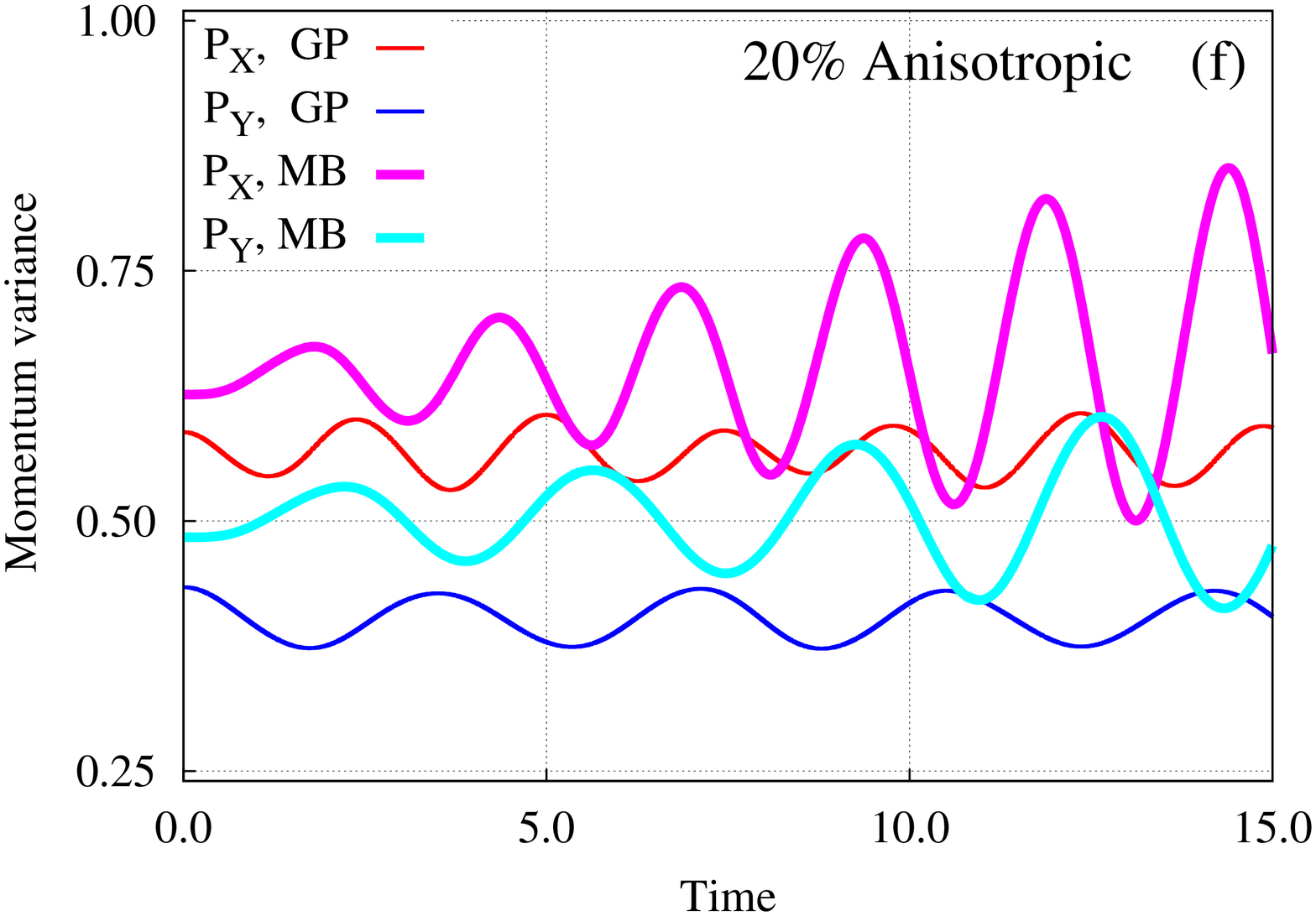}
\end{center}
\vglue 0.25 truecm
\caption{(Color online) 
Anisotropy in the breathing dynamics following an interaction quench.
Shown and compared are the many-body
results for $N=1000$ bosons 
using $M=3$ time-adaptive orbitals 
and the mean-field results (equivalent to $M=1$).
(a),(b),(c) The time-dependent many-particle position variance per particle,
$\frac{1}{N}\Delta^2_{\hat X}(t)$ and $\frac{1}{N}\Delta^2_{\hat Y}(t)$,
and (d),(e),(f) 
the momentum variance,
$\frac{1}{N}\Delta^2_{\hat P_X}(t)$ and $\frac{1}{N}\Delta^2_{\hat P_Y}(t)$,
following an interaction quench
from $\Lambda=\lambda_0(N-1)=0.09$ to $0.18$ at $t=0$
are plotted.
For the isotropic system, panels (a),(d), the quantities along the $y$ and $x$ directions are, of course, equal.
For the anisotropic systems, panels (b),(e) and (c),(f),
the time-dependent position variance 
in the $y$ and $x$ directions do not cross each other at the mean-field level,
and similarly the momentum quantities,
signifying that the position (momentum) density in the $y$ direction is always
wider (narrower) than that in the $x$ direction.
This, however, is no longer the case at the many-body level: 
A system whose position (momentum) density along the $y$ direction is wider (narrower)
than that along the $x$ direction can actually have a smaller (larger) position (momentum) variance
along this direction.
Note that the systems are essentially fully condensed, see Fig.~\ref{f5}.
See the text for further discussion.
The quantities shown are dimensionless.}
\label{f4}
\end{figure}

\begin{figure}[!]
\begin{center}
\hglue -1.0 truecm
\includegraphics[width=0.445\columnwidth,angle=-90]{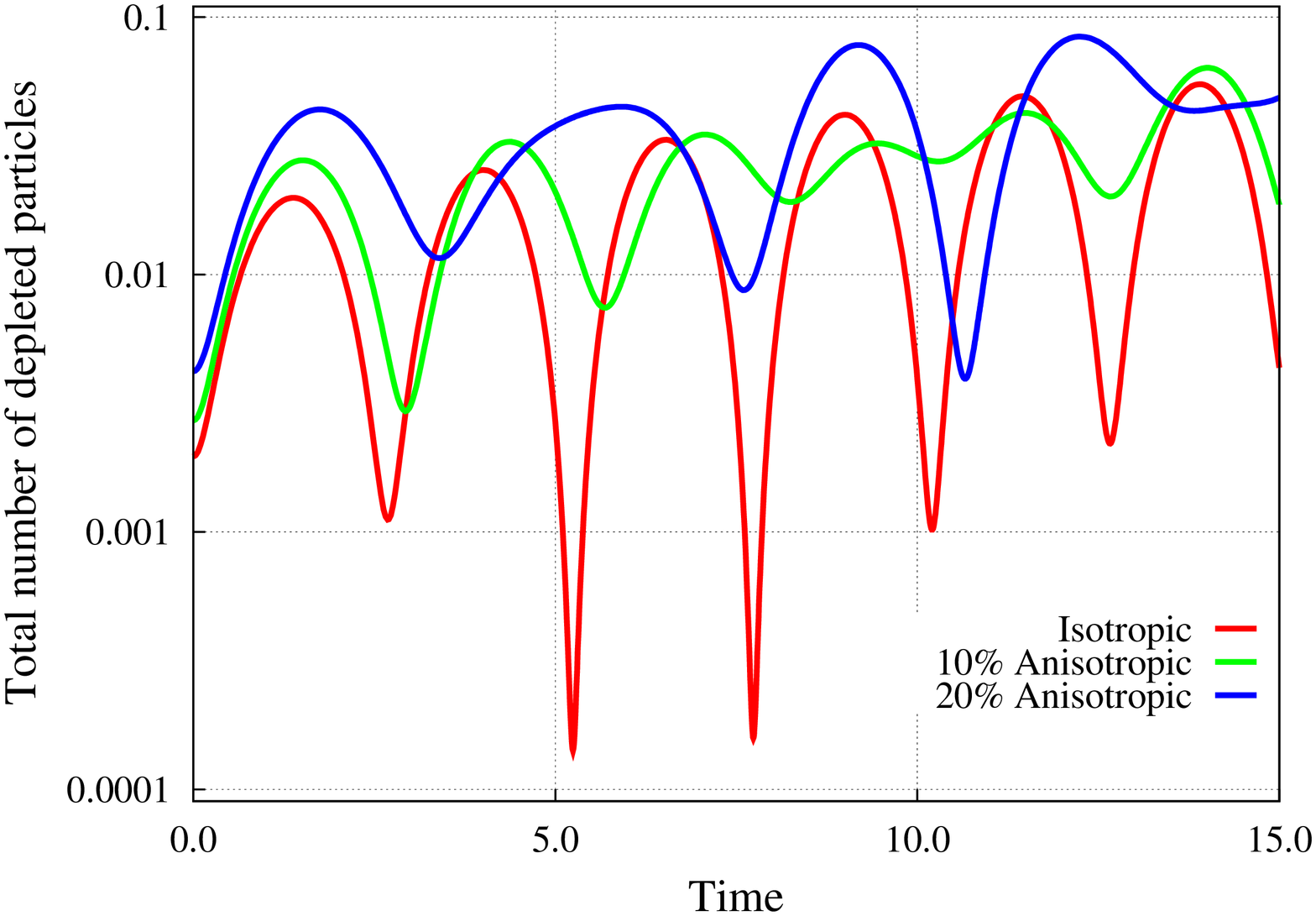}
\end{center}
\vglue 0.75 truecm
\caption{(Color online)
Depletion in the breathing dynamics following an interaction quench.
Shown is the total number of depleted particles outside the condensed mode of
$N=1000$ bosons, computed with $M=3$ time-adaptive orbitals,
in the out-of-equilibrium dynamics of Fig.~\ref{f4}.
Note the values on the y-axis. 
See the text for further discussion.
The quantities shown are dimensionless.}
\label{f5}
\end{figure}

At $t=0$ the interaction parameter is suddenly quenched to twice its value
$\Lambda'=\lambda'_0(N-1)=0.18$. 
We ask what the out-of-equilibrium dynamics of the system would be like.
Figs.~\ref{f4} and \ref{f5} collect the results.   
A complementary and comparative investigation for $N=10$ bosons 
and the same interaction-quench 
scenario is presented in the Appendix.

Following the quench of the interaction, 
the density performs breathing oscillations \cite{Bonz,Peter_2013,MCTDHB_3D_dyn},
i.e., its widths along the $y$ and $x$ directions varies in time.
Furthermore,
since the interaction is repulsive and at $t=0$ suddenly increased,
the density first expands at short times, i.e.,
the cloud initially broadens both along the $y$ and $x$ directions.

We have performed the calculations of the out-of-equilibrium dynamics
both at the Gross-Pitaevskii level ($M=1$) and
at the many-body level ($M=3$).
The two-dimensional time-dependent many-body density per particle, $\frac{\rho(\r;t)}{N}$,
essentially coincides
with the corresponding mean-field density.  
This is quite reasonable and expected 
due to the initial marginal depletion fraction [$\frac{n_1(0)}{N} > 0.999 \, 9(9)$]
which remains very small throughout the evolution of the BEC in time, see below.
As far as the condensed fraction is considered,
the system remains essentially fully condensed,
also see \cite{TD_Variance,TD_Variance_BEC} for respective
out-of-equilibrium studies in a one-dimensional trap.

In Fig.~\ref{f4} the time-dependent many-particle position variance per particle,
$\frac{1}{N}\Delta^2_{\hat X}(t)$ [panels (a), (b), and (c)],
and momentum variance,
$\frac{1}{N}\Delta^2_{\hat P}(t)$ [panels (d), (e), and (f)],
are shown.
The many-body and mean-field results are compared for each of the three trap anisotropies
$\beta=0\%$, $10\%$, and $20\%$.
In Fig.~\ref{f5} the total number of depleted particles outside 
the condensed mode [$\alpha_1(\r;t)$ natural orbital] 
are shown as a function of time. 
The systems are essentially fully condensed with only
a fraction of a single particle being depleted.

We start by analyzing the out-of-equilibrium dynamics in the isotropic trap, see Fig.~\ref{f4}a,d.
The results and discussion are to assist one in analyzing the dynamics in the anisotropic traps. 
For the isotropic trap, the quantities along the $y$ and $x$ directions, of course, coincide.
Both the many-body and mean-field variances vary in time in an oscillatory manner.
However, there are a couple of clearly visible differences.
The first, 
is that the mean-field variances oscillate with a rather constant amplitude,
whereas the many-body variances oscillates with a (slowly) growing amplitude.
The latter can be attributed to the (slowly) growing amount of depleted particles, see Fig.~\ref{f5},
albeit less than one tenth of a particle
is outside the condensed mode!
Indeed, the variance is a highly sensitive probe of correlations even when
the system is practically condensed \cite{Variance, TD_Variance}.
The second difference, 
is the opposite dynamical behavior of the variances at short times
when computed at the many-body and Gross-Pitaevskii levels.
Despite the expansion of the cloud at short times,
the time-dependent position variance increases and momentum variance decreases,
implying that the many-body contributions to the variance
$\Delta_{\hat x, MB}^2(t)=\Delta_{\hat y, MB}^2(t)$
and $\Delta_{\hat p_x, MB}^2(t)=\Delta_{\hat p_y, MB}^2(t)$
are opposite in sign with respect to and dominate the mean-field terms
$\Delta_{\hat x, density}^2(t)=\Delta_{\hat y, density}^2(t)$
and $\Delta_{\hat p_x, density}^2(t)=\Delta_{\hat p_y, density}^2(t)$.
This is an appealing time-dependent many-body effect
taking place in macroscopic Bose systems.

We now turn to the anisotropy of the variance in the $y$ and $x$ directions during the breathing dynamics,
see Fig.~\ref{f4}b,c,e,f. 
When the trap becomes anisotropic,
the time-dependent quantities along the $y$ direction `split' from the quantities
along the $x$ direction.
Since the isotropic trap is made anisotropic by a stretch
along the y-axis,
the `base line' of the
position variance along the $y$ direction is shifted up to higher values,
and the `base line' of the
momentum variance along
the $y$ axis is shifted down to lower values,
at least as far as the (initial conditions and) short-time dynamics is concerned.
Obviously, this shift is larger for $20\%$ anisotropy than for $10\%$ anisotropy,
compare Fig.~\ref{f4}b,e and Fig.~\ref{f4}c,f. 
All in all,
one can
anticipate from (the geometry of the trap and) 
the shape of the density 
the anisotropy of the time-dependent position and momentum variances,
at least for short times.
This result extends what is found in Subsec.~\ref{Ex1}
for the ground-state of an essentially fully-condensed system.
 
But the geometrical picture of the anisotropy of the variance 
emerging at short times 
changes in time.
At the mean-field level, the position variances in the $y$ and $x$ directions
(which are different in the anisotropic trap) oscillate with a rather constant amplitude.
Consequently, they do not cross each other,
indicating that during the breathing dynamics the
cloud's density remains wider along the
$y$ direction than along the $x$ direction. 
The corresponding momentum variances also oscillate with a rather constant amplitude
and, therefore, do not cross each other as well.
This implies that
during the dynamics the momentum density
of the cloud stays narrower along the $y$ direction than along the $x$ direction.

At the many-body level, on the other hand,
the variances along the $y$ and $x$ directions oscillate with a (slowly) growing amplitudes.
Therefore, at a certain point in time the $y$ and $x$ position variances
must cross each other for the first time.
Similarly, the $y$ and $x$ momentum variances must also cross at some point in time.
The smaller the anisotropy,
the earlier is this time, 
compare Fig.~\ref{f4}b,e for $\beta=10\%$
and Fig.~\ref{f4}c,f for $\beta=20\%$.
Thus,
the presence of even the slightest time-dependent depletion leads
to a large influence on the time-dependent variance at the many-body level:
The anisotropy of the variance for a whole time intervals 
is opposite to the anisotropy of the density.
Then, a simple analysis shows that
$\Delta_{\hat x, MB}^2(t)-\Delta_{\hat y, MB}^2(t) >
\Delta_{\hat y, density}^2(t)-\Delta_{\hat x, density}^2(t) > 0$
and 
$\Delta_{\hat p_y, MB}^2(t)-\Delta_{\hat p_x, MB}^2(t) >
\Delta_{\hat p_x, density}^2(t)-\Delta_{\hat p_y, density}^2(t) > 0$.

We have repeated the investigation of the out-of-equilibrium scenarios for a system
of $N=10$ bosons and the same interaction parameters.
The results are collected in the Appendix, see Fig.~\ref{f6} and \ref{f7}.
The similarity of the out-of-equilibrium 
results for the same interaction parameter and different numbers of particles
($N=1000$ bosons in the present subsection, $N=10$ in the Appendix),
together with analogous behavior in the dynamics of larger systems
in one-dimensional traps \cite{TD_Variance,TD_Variance_BEC},
provide, in our opinion, strong evidences
that the effects of the anisotropy of the time-dependent position and momentum variances
in essentially fully-condensed BECs 
persist in the limit of an infinite number of particle. 

All in all,
we have discussed two out-of-equilibrium effects associated with the variance
during the breathing dynamics of essentially-condensed
trapped bosons.
At short times, 
it is the decrease (increase) of the position (momentum) variance in contrast
to the increase (decrease) of the width of the position (momentum) density.
At longer times,
it is the opposite behavior of the anisotropy of the variances 
in position and momentum spaces
when computed at the mean-field and many-body levels. 

\section{Conclusions}\label{Conclusions}

We have investigated in the present work
the variance of the position and momentum many-particle operators
of structureless bosons
interacting by a long-range inter-particle interaction
and trapped in a two-dimensional single-well
anharmonic potential.
In the first investigation, 
that of the pathway from condensation to fragmentation of the ground state,
we find out that, although the density of the cloud is broader along the y-axis than along the x-axis,
the position variance can behave in an opposite manner, namely,
be larger along the x-axis than along the y-axis.
Similarly, the momentum variance can be larger along the y-axis than along the x-axis.
This opposite anisotropy of the variance with respect to the density 
is a counterintuitive many-body effect
that emerges when the ground-state of the 
bosonic system is fragmented.

In the second study, 
that of the out-of-equilibrium breathing dynamics of a BEC, 
we find out that, 
already when a fraction (even a tenth) of a boson is depleted,
qualitative differences
between the many-body and mean-field variances arise.
Explicitly, the time-dependent 
many-body position variance can show opposite
behavior of the anisotropy between the
y-axis and x-axis with respect to the mean-field quantities,
despite the system being essentially fully condensed.
Corresponding
results hold for the time-dependent 
many-body momentum variance in comparison with the mean-field quantities.

Both the ground-state and out-of-equilibrium scenarios 
suggest a wealth of effects emanating from the many-body
term of the variance both in position and momentum spaces in interacting 
trapped many-boson systems,
in two spatial dimensions.
The anisotropy of the variance
advocates that it can be used 
to characterize the
strength of correlations 
along the $y$ and $x$ directions
in the system.
We have seen that 
such many-body effects encoded within the variance do not necessarily
match with the information that can
be extracted based on the shape of system's density.
This is in sharp
contrast to the text-book example of a single particle
discussed in the introduction.

\section*{Acknowledgements}

This paper is dedicated to Professor Hans-Dieter Meyer, a dear colleague and friend, on
the occasion of his 70th birthday.
This research was supported by the Israel Science Foundation (Grant No. 600/15).
Partial financial support by the
Deutsche Forschungsgemeinschaft (DFG) is acknowledged.
We thank Kaspar Sakmann for discussions. 
Computation time on the Cray XC40 
system Hazelhen at the High Performance Computing Center
Stuttgart (HLRS) is gratefully acknowledged.

\appendix

\section{Further computational details and convergence}\label{APP_Num}

\begin{figure}[!]
\hglue -1.0 truecm
\includegraphics[width=0.345\columnwidth,angle=-90]{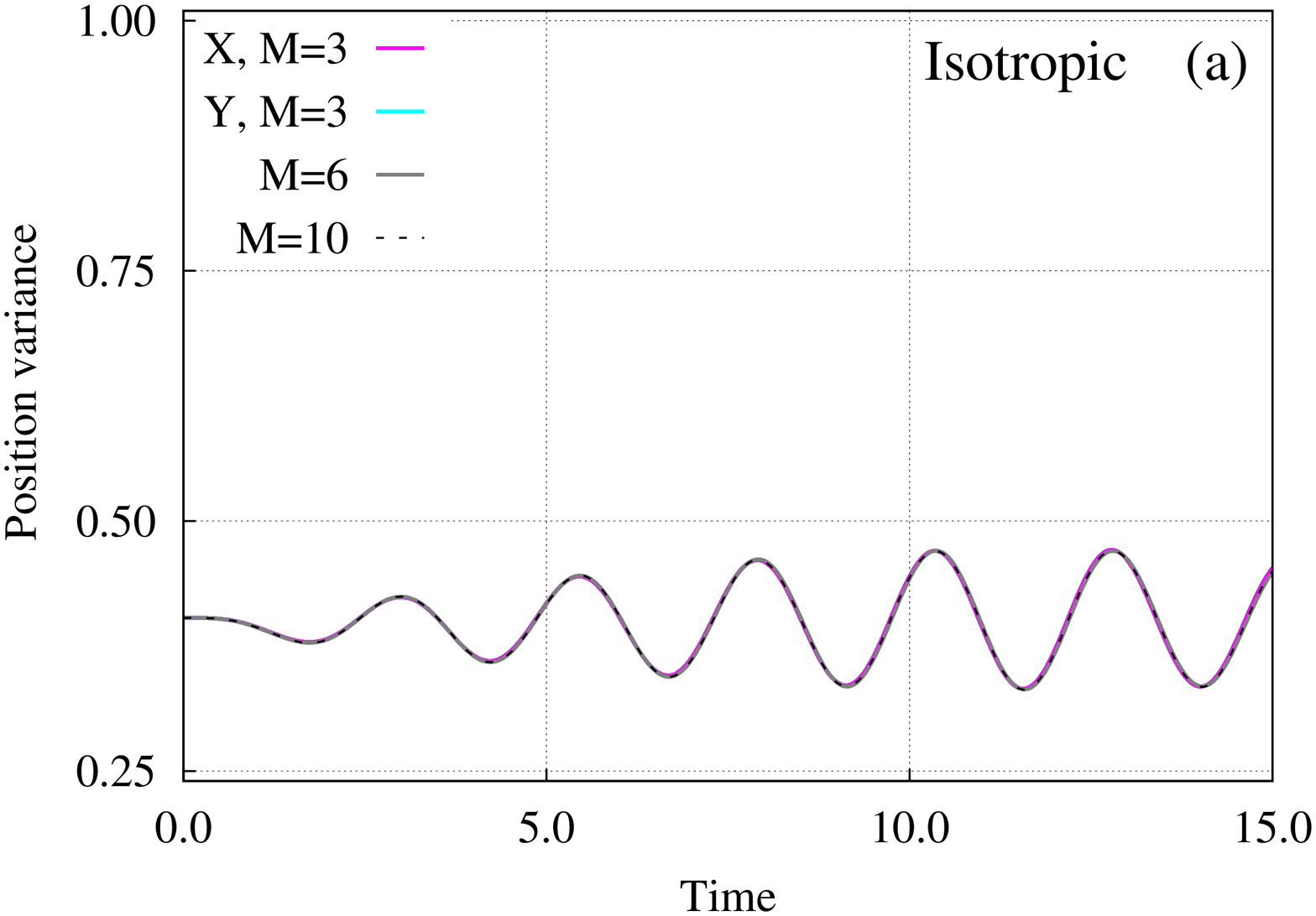}
\includegraphics[width=0.345\columnwidth,angle=-90]{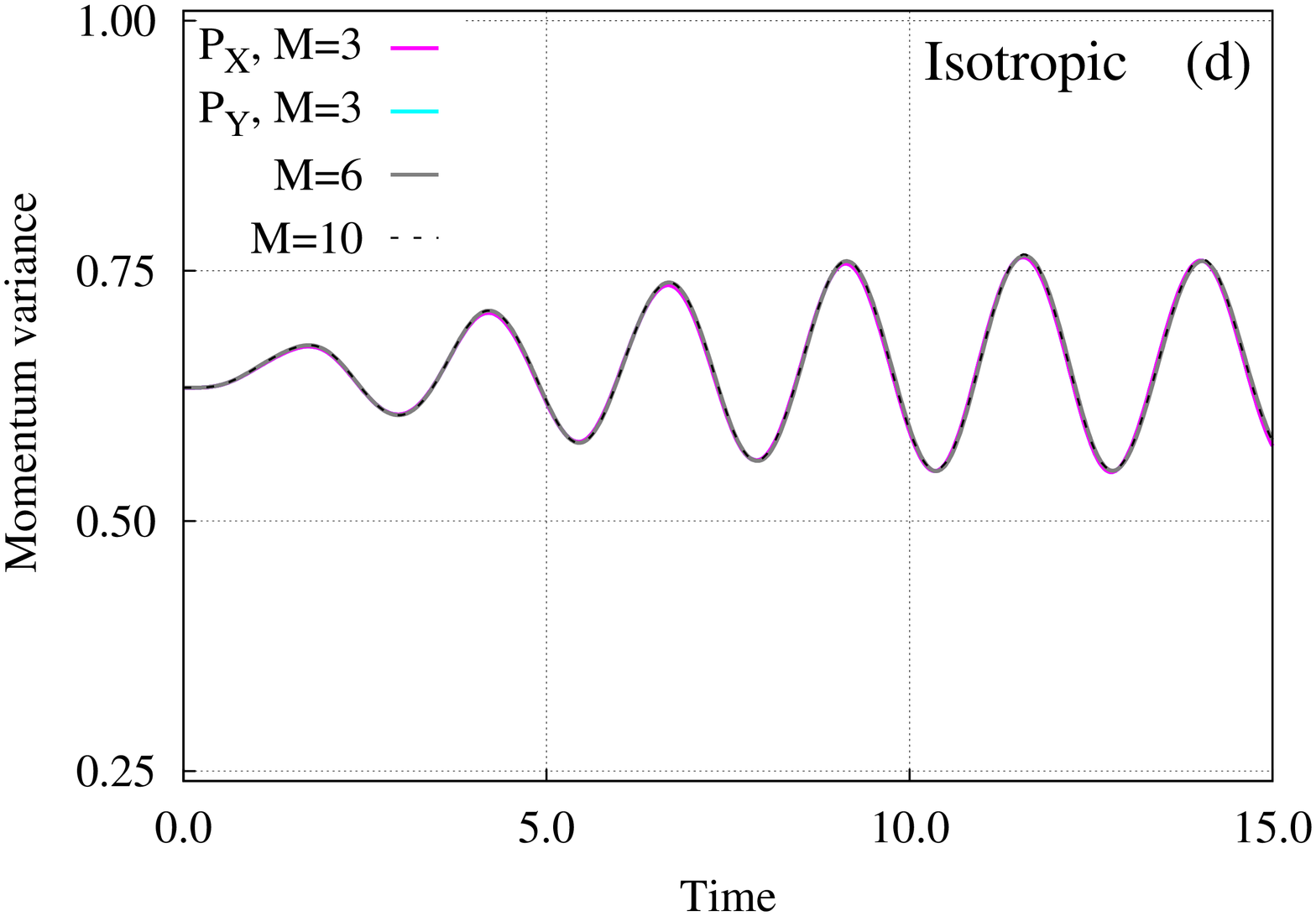}
\hglue -1.0 truecm
\includegraphics[width=0.345\columnwidth,angle=-90]{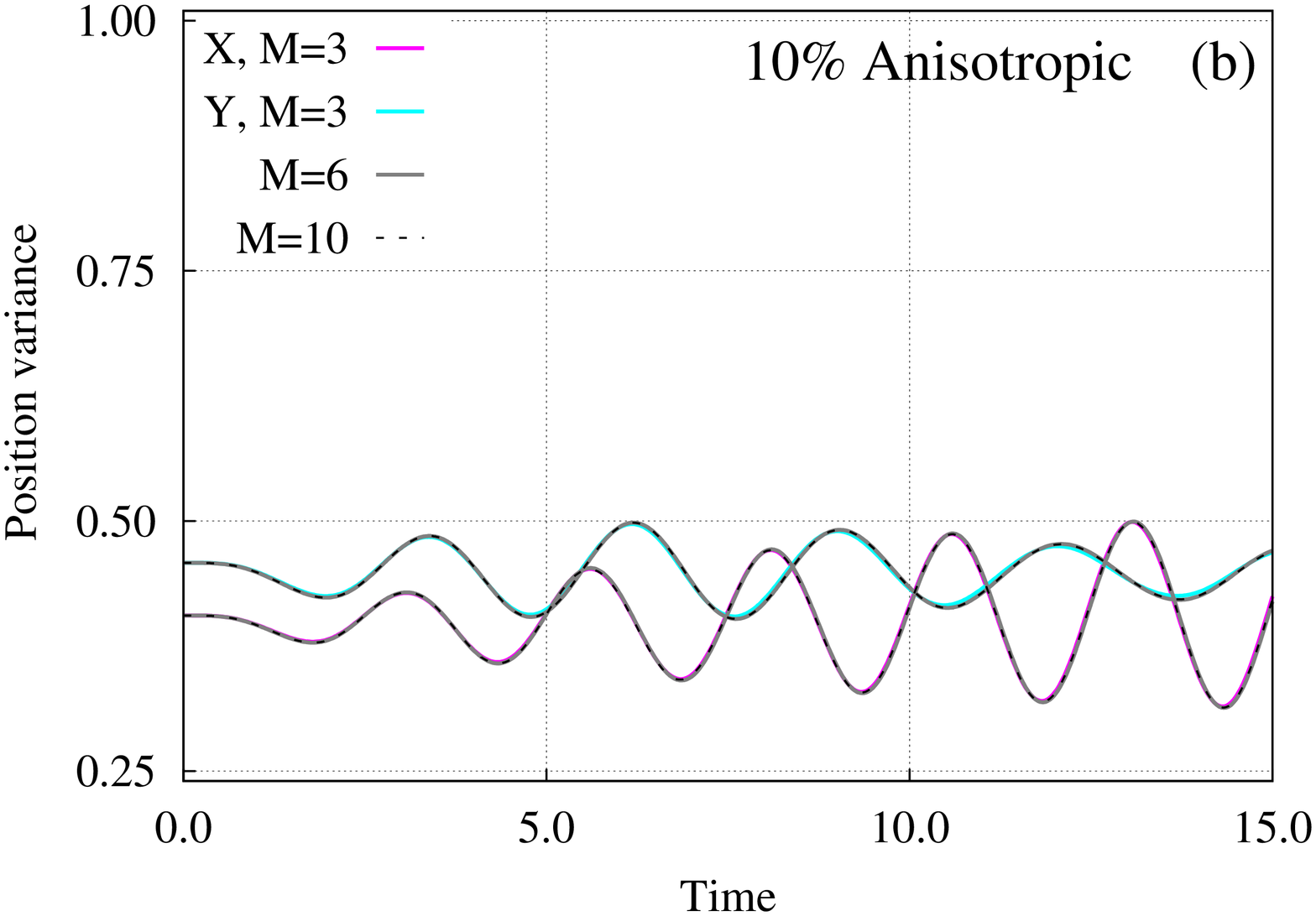}
\includegraphics[width=0.345\columnwidth,angle=-90]{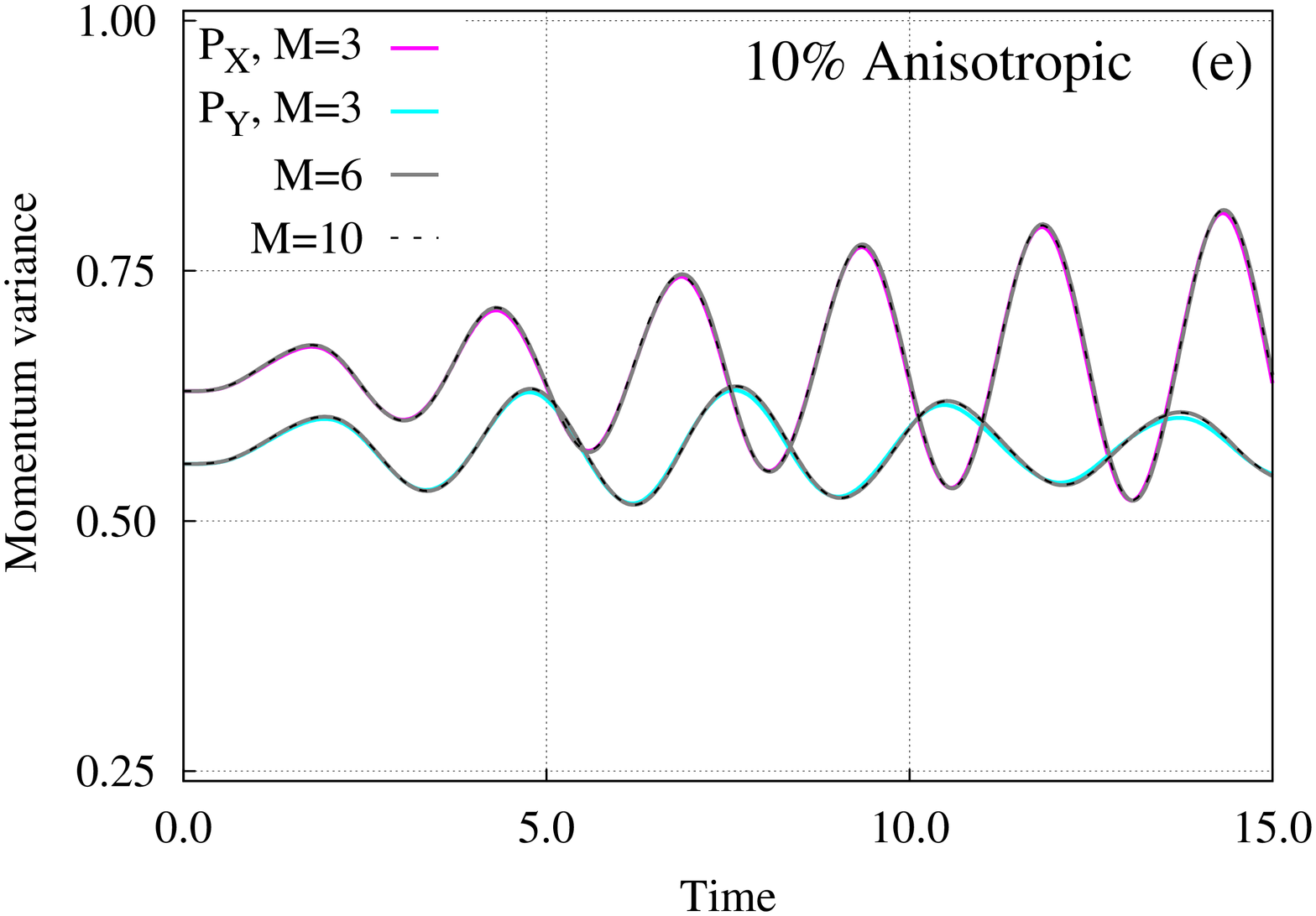}
\hglue -1.0 truecm
\includegraphics[width=0.345\columnwidth,angle=-90]{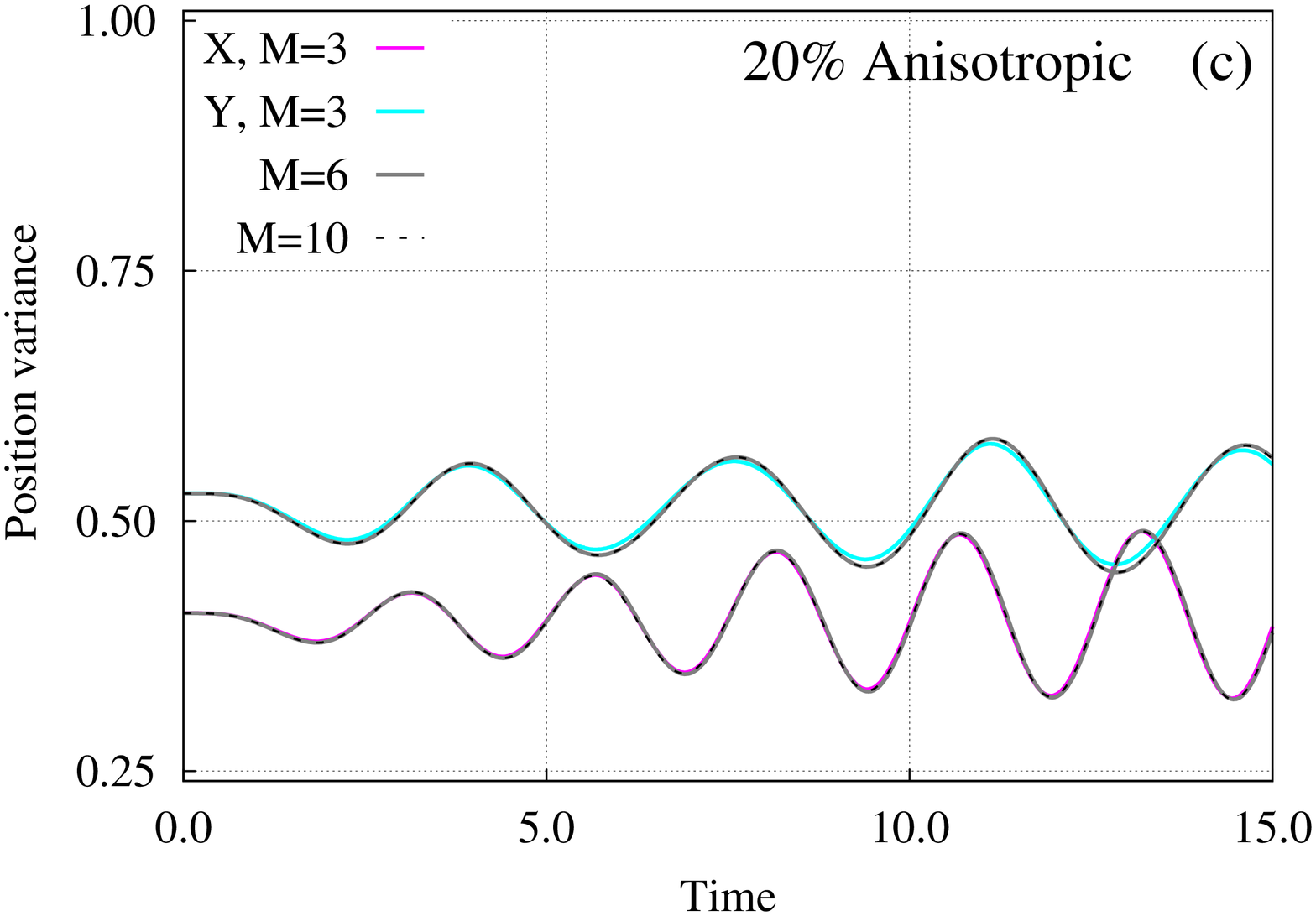}
\includegraphics[width=0.345\columnwidth,angle=-90]{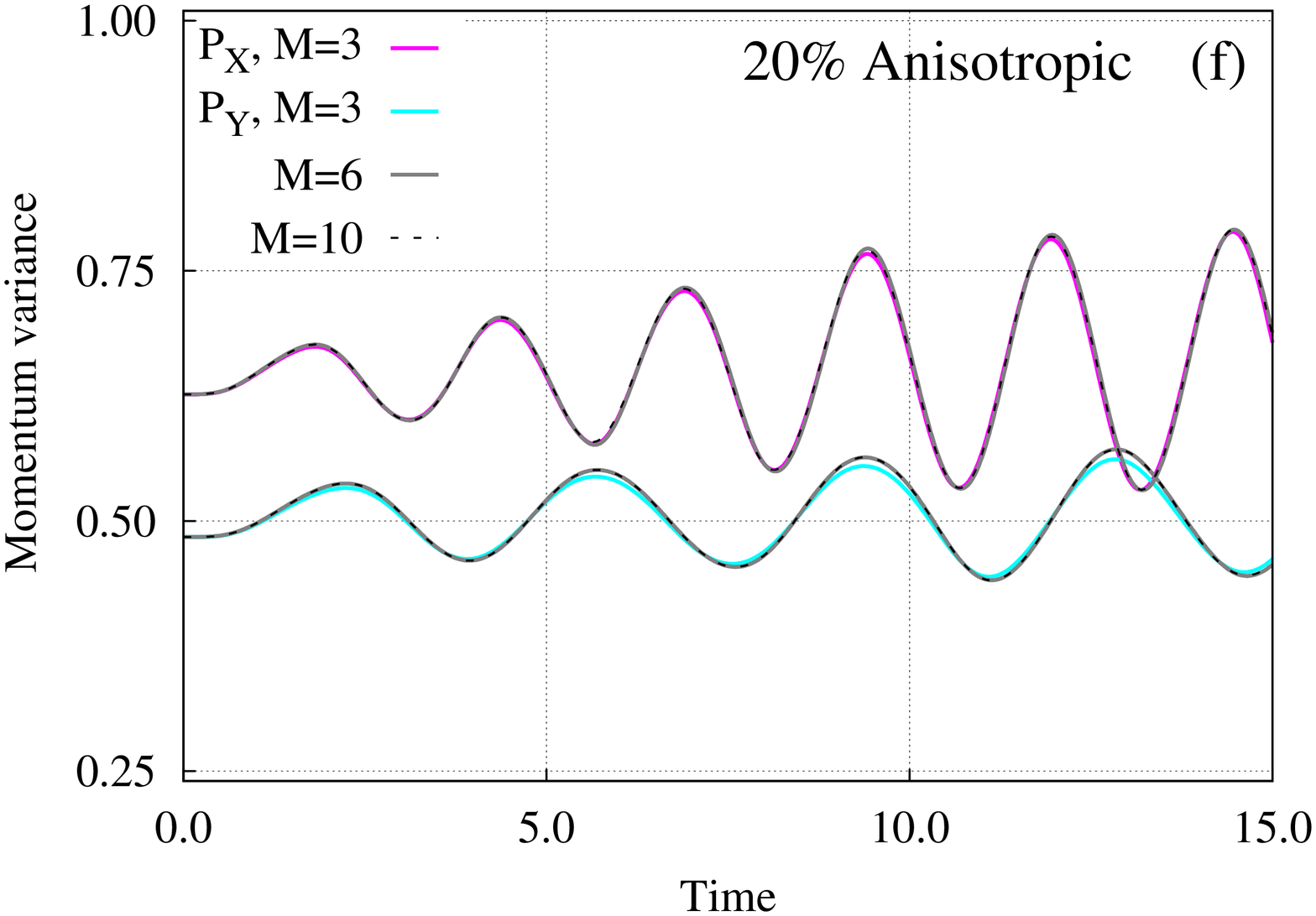}
\vglue 0.75 truecm
\caption{(Color online)
Convergence of the time-dependent many-particle position variance per particle,
$\frac{1}{N}\Delta^2_{\hat X}(t)$ and $\frac{1}{N}\Delta^2_{\hat Y}(t)$ [panels (a), (b), and (c)],
and the momentum variance,
$\frac{1}{N}\Delta^2_{\hat P_X}(t)$ and $\frac{1}{N}\Delta^2_{\hat P_Y}(t)$ [panels (d), (e), and (d)],
with the number of time-adaptive orbitals $M$ used in the MCTDHB computations 
for systems consisting of $N=10$ bosons in traps of anisotropies $\beta=0\%$, $10\%$, and $20\%$
following an interaction quench
from $\Lambda=\lambda_0(N-1)=0.09$ to $0.18$ at $t=0$.
Compare to Fig.~\ref{f4}.
It is found that the results with $M=3$ accurately describe the physics and
the results with $M=6$ and $M=10$ orbitals lie atop each other. 
The quantities shown are dimensionless.}
\label{f6}
\end{figure}

\begin{figure}[!]
\begin{center}
\hglue -1.0 truecm
\includegraphics[width=0.445\columnwidth,angle=-90]{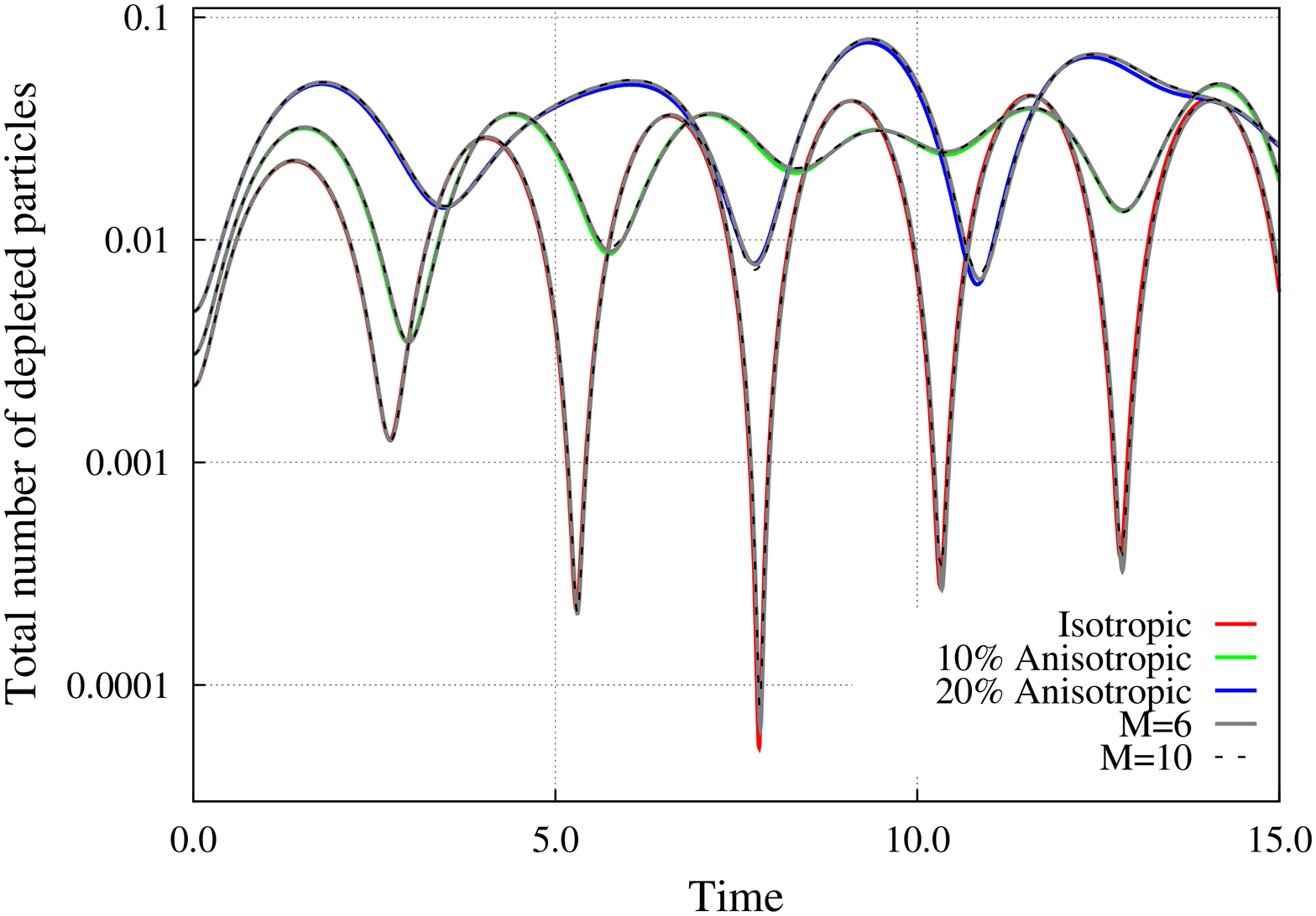}
\end{center}
\vglue 0.75 truecm
\caption{(Color online)
Convergence of the total number of depleted particles 
with the number of time-adaptive orbitals $M$ used in the MCTDHB computations 
for systems consisting of $N=10$ bosons
following an interaction quench
from $\Lambda=\lambda_0(N-1)=0.09$ to $0.18$ at $t=0$, see Fig.~\ref{f6}.
It is found that the results with $M=3$ accurately describe the physics and
the results with $M=6$ and $M=10$ orbitals lie atop each other. 
Compare to Fig.~\ref{f5}.
The quantities shown are dimensionless.}
\label{f7}
\end{figure}

The multiconfigurational time-dependent Hartree for bosons 
(MCTDHB) method \cite{MCTDHB1,MCTDHB2,Kaspar_The,book_nick,Axel_The,book_MCTDH} 
is used in the present work to compute the
ground-state and out-of-equilibrium properties
of trapped bosons in two spatial dimensions 
interacting by a long-range inter-particle interaction.
The maximal configurational space are $501\,501$
for $N=1000$ bosons and $M=3$ orbitals
and $1\,961\,256$
for $N=10$ bosons and $M=15$ orbitals.  
We use the numerical implementation in the software packages \cite{MCTDHB_LAB,package}.
To obtain the ground state
we propagate the MCTDHB equations of motion in imaginary time \cite{Benchmarks,MCHB}. 
For the computations the many-body Hamiltonian is represented 
by $128^2$ exponential discrete-variable-representation grid points
(using a Fast-Fourier Transform routine) in a box of size $[-10,10) \times [-10,10)$.
Convergence of the occupation numbers (depletion) and the position and momentum 
variance with increasing number of `filled shells', 
namely, $M=3$, $6$, $10$, and $15$ time-adaptive orbitals, 
is demonstrated for $N=10$ bosons in Figs.~\ref{f2} and \ref{f3} for the ground state \cite{Variance}
and in Figs.~\ref{f6} and \ref{f7} for the out-of-equilibrium breathing dynamics \cite{TD_Variance},
also see in this context \cite{Brand_Cos}.
It is found that the results 
are nicely converged.

\end{document}